\documentclass[prd,preprintnumbers,superscriptaddress,nofootinbib,floatfix,onecolumn,notitlepage]{revtex4-1}

\usepackage{graphicx,epsfig,psfrag,amssymb,hyperref}
\usepackage{multirow}
\usepackage{color,graphicx,epsfig,psfrag,amsmath,empheq}
\usepackage{bm}
\usepackage{mathrsfs,amsfonts,, color}
\usepackage[caption=false]{subfig}
\usepackage{hepunits}
\usepackage{color}
\definecolor{darkgreen}{rgb}{0,0.5,0}

\newcommand{\Ref}[1]{Ref.~\cite{#1}}
\newcommand{\Refs}[1]{Refs.~\cite{#1}}
\newcommand{\Fig}[1]{Fig.~\ref{#1}}
\newcommand{\Figs}[2]{Figs.~\ref{#1} and \ref{#2}}
\newcommand{\Tab}[1]{Table~\ref{#1}}

\newcommand{\Sec}[1]{Sec.~\ref{#1}}
\newcommand{\Secs}[2]{Secs.~\ref{#1} and \ref{#2}}
\newcommand{\App}[1]{App.~\ref{#1}}
\newcommand{\Apps}[2]{Apps.~\ref{#1} and \ref{#2}}
\newcommand{\Eq}[1]{Eq.~(\ref{#1})}

\newcommand{\be}{\begin{eqnarray}}
\newcommand{\ee}{\end{eqnarray}}
\def\lsim{\mathrel{\rlap{\lower4pt\hbox{\hskip 0.5 pt$\sim$}}
\raise1pt\hbox{$<$}}}  
\newcommand{\dae}{DAE$\delta$ALUS}
\newcommand{\apr}{A^\prime}

\newcommand{\pizero}{\pi^0}

\newcommand{\Br}{\textrm{Br}}
\newcommand{\alphaEM}{\alpha_{\rm{EM}}}

\begin{document}

\title{\dae\ and Dark Matter Detection}

\author{Yonatan Kahn}
\email{ykahn@mit.edu}
\affiliation{Center for Theoretical Physics, Massachusetts Institute of Technology, Cambridge, MA 02139, U.S.A.}
\author{Gordan Krnjaic}
\email{gkrnjaic@perimeterinstitute.ca}
\affiliation{Perimeter Institute for Theoretical Physics, Waterloo, Ontario N2L 2Y5, Canada}
\author{Jesse Thaler}
\email{jthaler@mit.edu}
\affiliation{Center for Theoretical Physics, Massachusetts Institute of Technology, Cambridge, MA 02139, U.S.A.}
\author{Matthew Toups}
\email{mtoups@mit.edu}
\affiliation{Laboratory for Nuclear Science, Massachusetts Institute of Technology, Cambridge, MA 02139, U.S.A.}
\affiliation{Fermilab, Batavia, IL 60510, U.S.A.}

\begin{abstract}
Among laboratory probes of dark matter, fixed-target neutrino experiments are particularly well-suited to search for light weakly-coupled dark sectors.  In this paper, we show that the \dae\ source setup---an 800 \MeV\ proton beam impinging on a target of graphite and copper---can improve the present LSND bound on dark photon models by an order of magnitude over much of the accessible parameter space for light dark matter when paired with a suitable neutrino detector such as LENA. Interestingly, both \dae\ and LSND are sensitive to dark matter produced from off-shell dark photons. We show for the first time that LSND can be competitive with searches for visible dark photon decays, and that fixed-target experiments have sensitivity to a much larger range of heavy dark photon masses than previously thought. We review the mechanism for dark matter production and detection through a dark photon mediator, discuss the beam-off and beam-on backgrounds, and present the sensitivity in dark photon kinetic mixing for both the \dae/LENA setup and LSND in both the on- and off-shell regimes.
\end{abstract}

\preprint{MIT--CTP 4591}

\maketitle

\section{Introduction}
\label{sec:Introduction}

The gravitational evidence for dark matter (DM) is overwhelming \cite{Agashe:2014kda,Ade:2013zuv}, but most realistic DM scenarios predict some kind of non-gravitational interactions between DM and ordinary matter.  One ubiquitous prediction is that DM should have non-zero scattering cross sections off nuclei, which is the mechanism by which direct detection experiments search for DM in the galactic halo \cite{Goodman:1984dc, Lewin:1995rx}.  DM can also be produced in laboratory experiments, either at high energies at machines like the Large Hadron Collider \cite{Askew:2014kqa}, or at low energies through bremsstrahlung or rare hadron decays (see \Ref{Essig:2013lka} for a review). This low energy mode has been exploited to use fixed-target neutrino experiments such as LSND \cite{Athanassopoulos:1996ds} and MiniBooNE \cite{AguilarArevalo:2010cx} as production and detection experiments for sub-\GeV \ DM \cite{Batell:2009di, deNiverville:2011it, deNiverville:2012ij}, and it has been recently proposed to use the main injector beam at Fermilab paired with the NO$\nu$A detector \cite{Ayres:2007tu} to search for GeV-scale DM \cite{Dobrescu:2014ita}.\footnote{As of this writing, MiniBooNE is currently analyzing data taken in off-target mode for a dark sector search. The expanded off-shell reach we discuss in this paper could have important consequences for this search.}  A similar logic applies to electron beam fixed-target experiments \cite{Izaguirre:2013uxa,Izaguirre:2014dua,Diamond:2013oda}. 

\begin{figure}[t!] 
 \vspace{0.1cm}
\includegraphics[width=16cm]{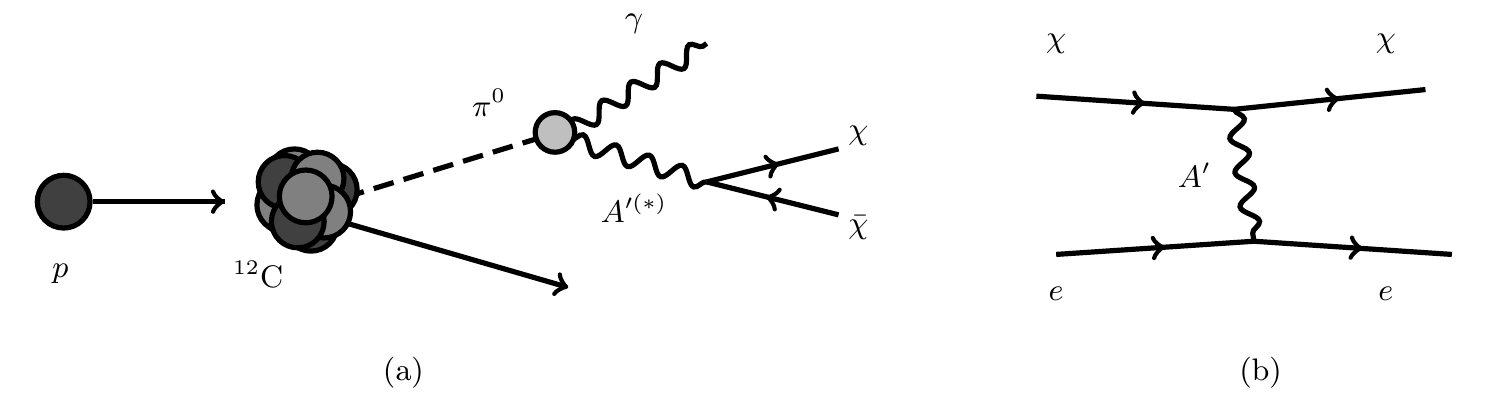}
  \caption{
{\bf Left (a):} schematic  diagram of DM production in proton-carbon collisions, through on- or off-shell dark photons $A'$ from exotic $\pi^0$ decays. {\bf Right (b):} DM scattering at a detector through the same dark photon $A'$. We focus on electron scattering in this paper, but the detector target may be protons or nuclei in alternative experimental setups.}
\label{fig:production}
\vspace{0cm}
\end{figure}

In this paper, we propose conducting a DM search using \dae\ \cite{Adelmann:2013isa} in close proximity to a large-volume neutrino detector such as the proposed LENA detector \cite{Wurm:2011zn}.\footnote{The study in  \Ref{Izaguirre:2014cza} also considers an underground accelerator paired with a large neutrino detector to search for light scalars of relevance to the proton radius puzzle. } \dae\ uses cyclotrons (peak power 8 MW, average power 1--2 MW) to produce a high-intensity 800 \MeV\ proton beam incident on a graphite and copper target (1 m of graphite liner inside a 3.75 m copper beam stop), creating a decay-at-rest neutrino source from stopped charged pions.  Proton-carbon scattering is also a rich source of neutral pions, and in scenarios involving a light weakly-coupled dark sector, rare $\pi^0$ decays to an on-shell dark mediator $A'$ can produce pairs of DM particles $\chi \overline{\chi} $ when $2 m_{\chi} < m_{\pi^0}$.  These DM particles can then be detected through neutral-current-like scattering in detectors designed to observe neutrinos, as illustrated in \Figs{fig:production}{fig:LENAGeometries}. A similar setup was the basis for existing LSND bounds on light DM \cite{Batell:2009di, deNiverville:2011it}, but we find that for light $\chi$, \dae\ can improve the reach of LSND by an order of magnitude in the visible-dark sector coupling $\epsilon^2$ after only one year of running.  This DM search is therefore an important physics opportunity for the initial single-cyclotron phase of \dae. 

We also find that both \dae\ and LSND are sensitive to DM production through \emph{off-shell} mediators in two distinct regimes, a fact that has been overlooked in the literature. Surprisingly, in the lower regime ($m_{A'} < 2m_\chi$), sensitivity to an off-shell $A'$ can be superior compared to a heavier on-shell $A'$. In the upper regime ($m_{A'} > m_{\pizero}$), existing LSND limits are considerably stronger than previously reported, and the \dae\ sensitivity can extend up to $m_{A'} \simeq 800 \ \MeV$ rather than cutting off at $m_{A'} \simeq m_{\pizero}$. Indeed, the observation that DM produced from meson decays can probe $A'$ masses much heavier than the meson mass expands the sensitivity of the entire experimental
 program to discover DM in proton-beam fixed-target searches. As \Figs{fig:money}{fig:g2favored} illustrate, the combination of updated LSND bounds and projected  \dae\ sensitivity covers a broad range of DM and mediator masses, and is even competitive with searches for visibly-decaying mediators in certain regions of parameter space.

The search strategies for MeV-scale DM at both \dae\ and LSND are very similar, so it is worth pointing out the potential advantages of \dae\ compared to LSND: 
\begin{itemize}
\item \emph{Higher energy range}. The LSND $\nu_e - e^-$ elastic scattering measurement \cite{Auerbach:2001wg}, which has been used to set limits on light DM, focused on the recoil electron energy range $E_e \in [18,52] \ \MeV$, where a \u{C}erenkov detector can use directionality to discriminate against decay-at-rest neutrino backgrounds. This strategy is optimal for a heavier DM search ($m_{\chi} \gtrsim 40~\MeV$) where the kinetic energy available for scattering is smaller. Here, we propose a search with \dae/LENA in the higher energy range $E_e \in [106,400] \ \MeV$, well above the thresholds from decay-at-rest backgrounds, which is optimal for lighter DM ($m_{\chi} \lesssim 20~\MeV$). The specialized target at \dae, designed to reduce the decay-in-flight component of the neutrino beam, makes such a high-energy search possible by reducing decay-in-flight backgrounds.\footnote{In principle, LSND could have done such a high-energy search as well. It may be possible to derive stronger limits than those from the LSND electron scattering measurement by using LSND's measurement of $\nu_e \, \textrm{C} \to e^- \, X$ at 60--200 MeV \cite{Athanassopoulos:1997er}.}
\item \emph{Higher luminosity}.  A single \dae\ cyclotron with a 25\% duty cycle and peak power 8 MW can deliver $4.9 \times 10^{23}$ protons on target per year, producing $7.5 \times 10^{22}$ $\pi^0$ per year, compared to $10^{22}$ $\pi^0$ over the life of the LSND experiment. 
\item \emph{Larger acceptance}.  At LSND, the source was placed a distance of 30 m from the neutrino detector, whereas the \dae\ source can be placed as close as 20 m to the detector, increasing the angular acceptance for DM scattering.  In addition, the detector length of LSND was 8.3 m, whereas \dae\ can be paired with a large neutrino detector like LENA in a geometry where the average path length through the detector is closer to 21~m, and the maximum path length is over 100~m.
\end{itemize}
Because we consider a dedicated DM search with \dae, we will optimize our cuts for each point in the dark sector parameter space. We will show that under conservative assumptions, a light DM search at \dae/LENA is systematics dominated. In particular, the improvements compared to LSND come almost exclusively from the optimized cuts rather than the higher luminosity and larger acceptance, though that conclusion could change with relatively modest improvements to the systematic uncertainties of neutrino-nucleon scattering cross sections.

The full \dae\ program \cite{Conrad:2010eu} includes multiple cyclotron-based neutrino sources placed at three different distances from a single detector such as LENA.  Because the earliest phase of \dae\ involves just a single ``near'' cyclotron-based neutrino source, we focus on pairing this neutrino source with a neutrino detector to perform a dedicated DM search.\footnote{One could also pair \dae\ with the proposed JUNO~\cite{Li:2013zyd}, Hyper-K~\cite{Abe:2011ts}, or water-based liquid scintillator~\cite{Alonso:2014fwf} detectors. While it may be possible to use an existing neutrino detector such as NO$\nu$A, beam-off backgrounds for an above-ground detector appear prohibitive.} For studies of other physics opportunities with a near cyclotron, see \Refs{Anderson:2011bi, Agarwalla:2011qf, Anderson:2012pn}.

% -----------------------------------------------------------------------------------------------------------------------------------------------------------------
%								Production and Scattering Diagrams
% -----------------------------------------------------------------------------------------------------------------------------------------------------------------

% -----------------------------------------------------------------------------------------------------------------------------------------------------------------
% LENA geometries
% -----------------------------------------------------------------------------------------------------------------------------------------------------------------

\begin{figure}[t!] 
\vspace{0.1cm}
\hspace{-0.5cm}
\includegraphics[width=17cm]{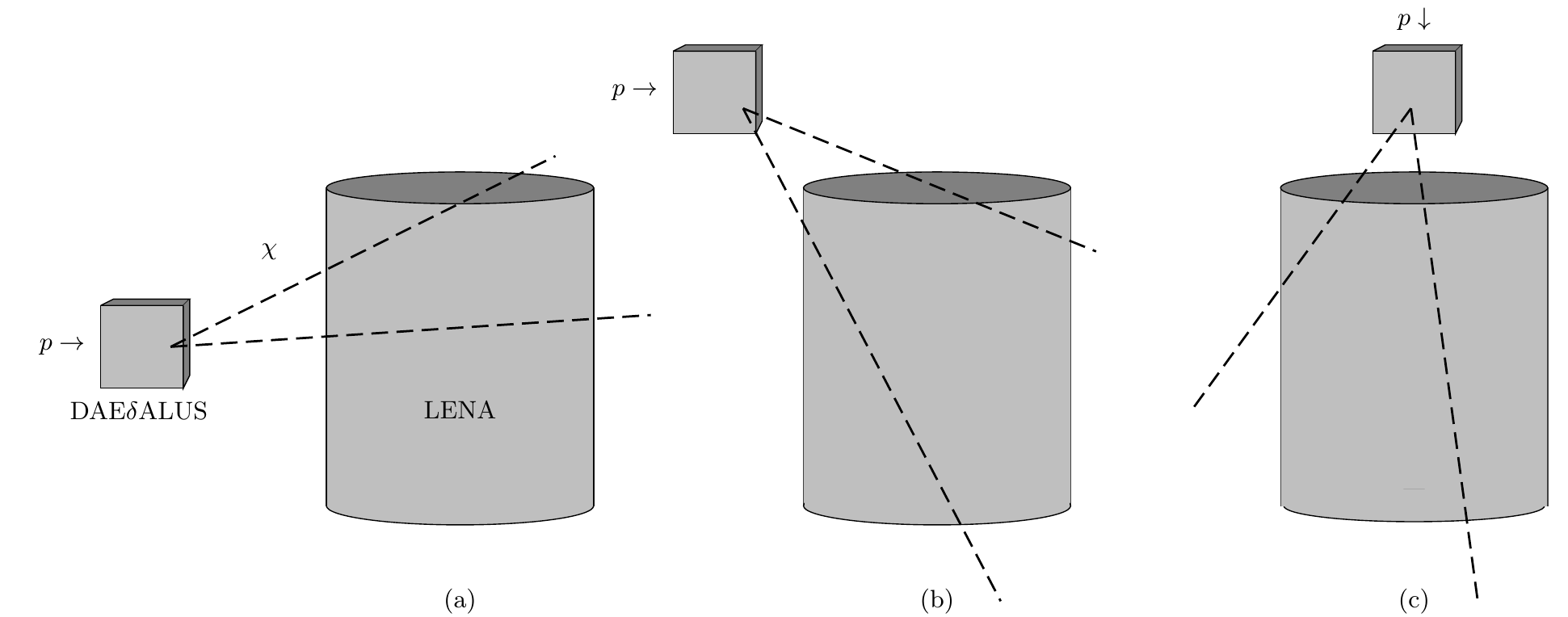}  ~~~
  \caption{Example \dae\ placements in the vicinity of the cylindrical LENA detector: midpoint (a), oblique (b), and on-axis (c). The dotted lines show some representative paths of $\chi$ through the detector volume. The projected yields for each configuration are displayed in \Fig{fig:ChangeGeo}. Note that for our sensitivity projections, we assume the DM incidence angle is always defined with respect to the incident proton direction.}
   \label{fig:LENAGeometries}
\vspace{0cm}
\end{figure}
% -----------------------------------------------------------------------------------------------------------------------------------------------------------------

% -----------------------------------------------------------------------------------------------------------------------------------------------------------------
% 								Money Plot
% -----------------------------------------------------------------------------------------------------------------------------------------------------------------

\begin{figure}[t!] 
 \hspace{-0.5cm}
   \vspace{-0.7cm}

\subfloat[]{\label{fig:money:a} \includegraphics[width=7.45cm]{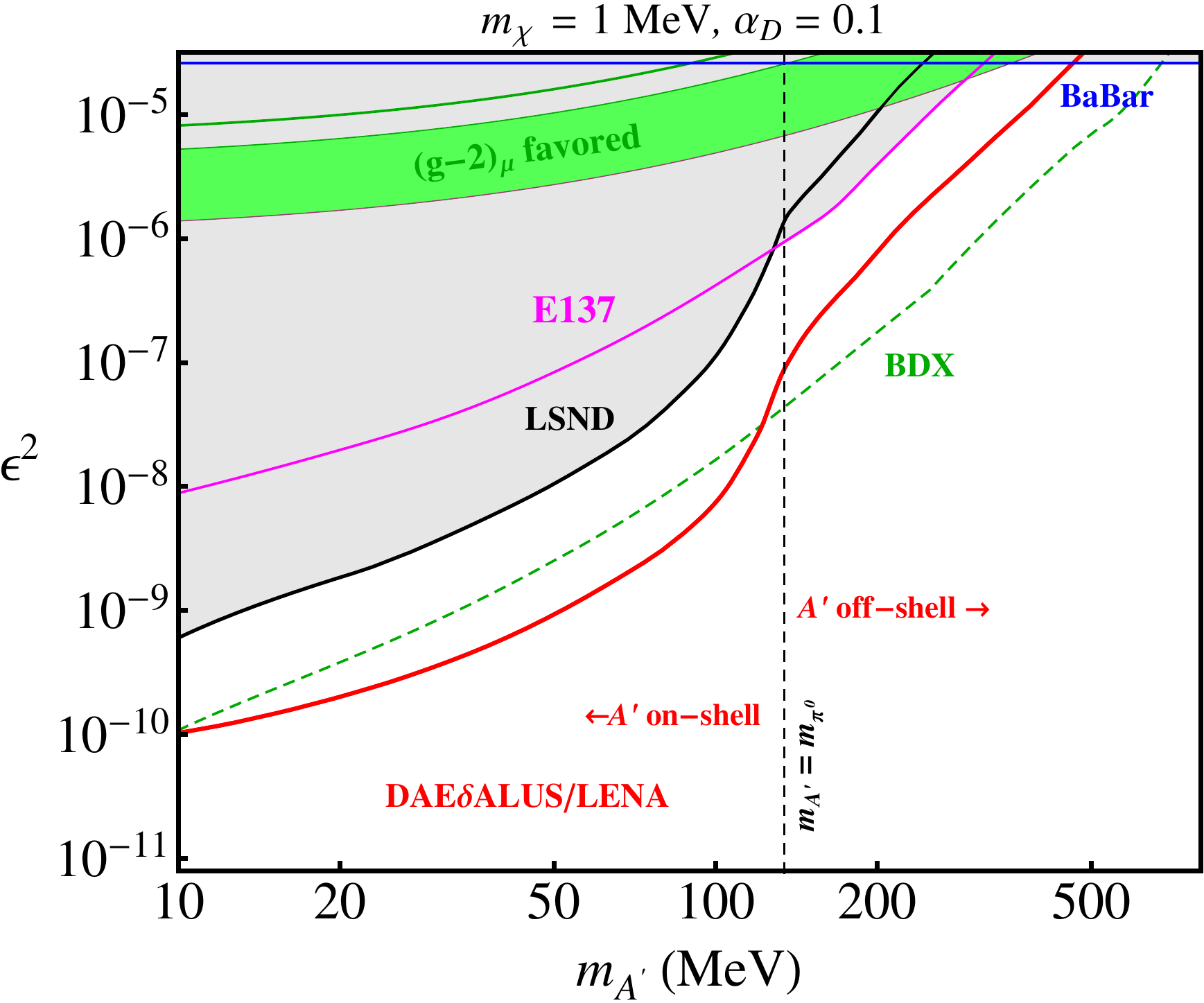}} ~~~~~~~~
 \subfloat[]{ \label{fig:money:b}  \includegraphics[width=7.45cm]{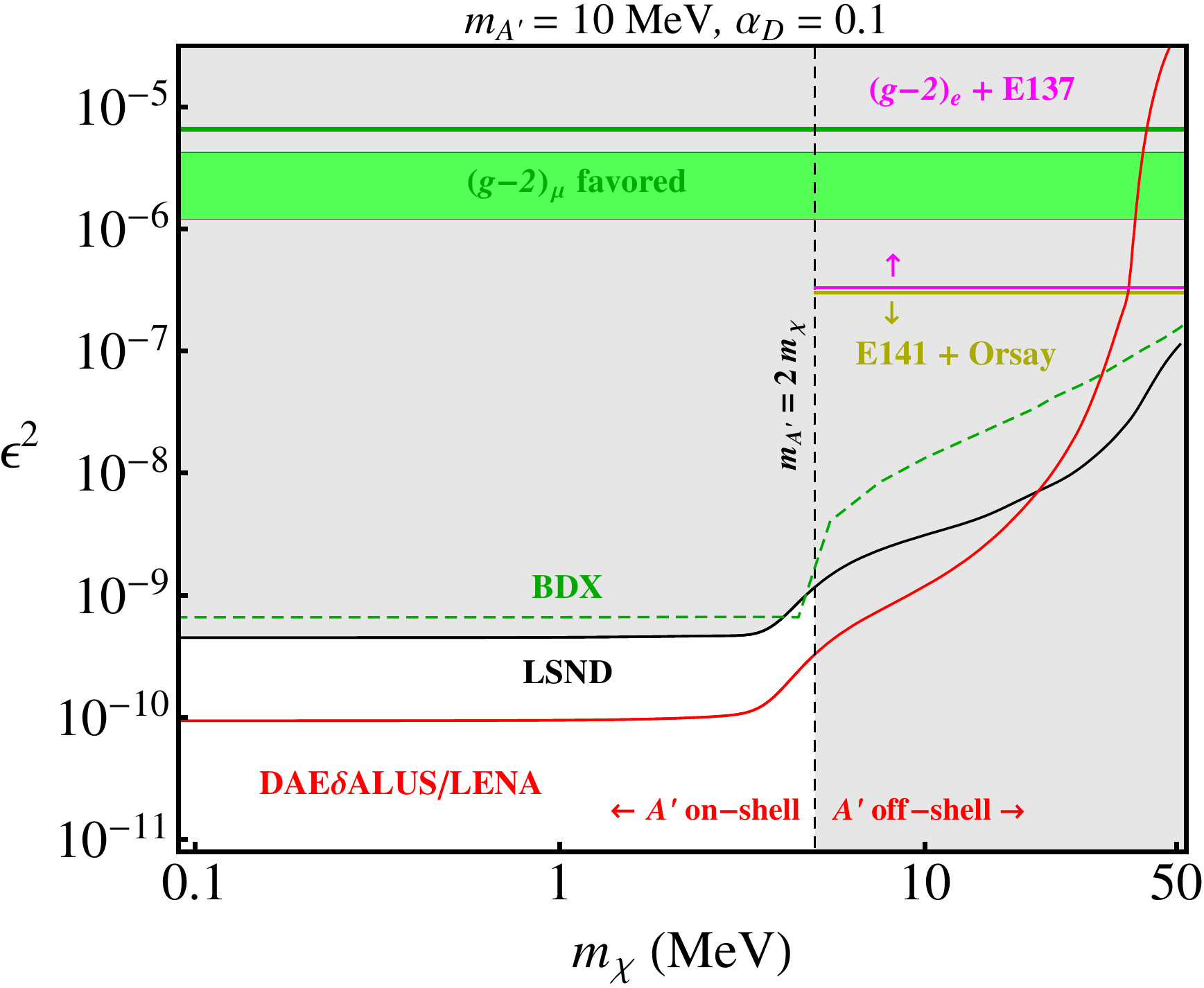}} \\ 
    \vspace{-0.4cm}
\subfloat[]{  \label{fig:money:c}\includegraphics[width=7.45cm]{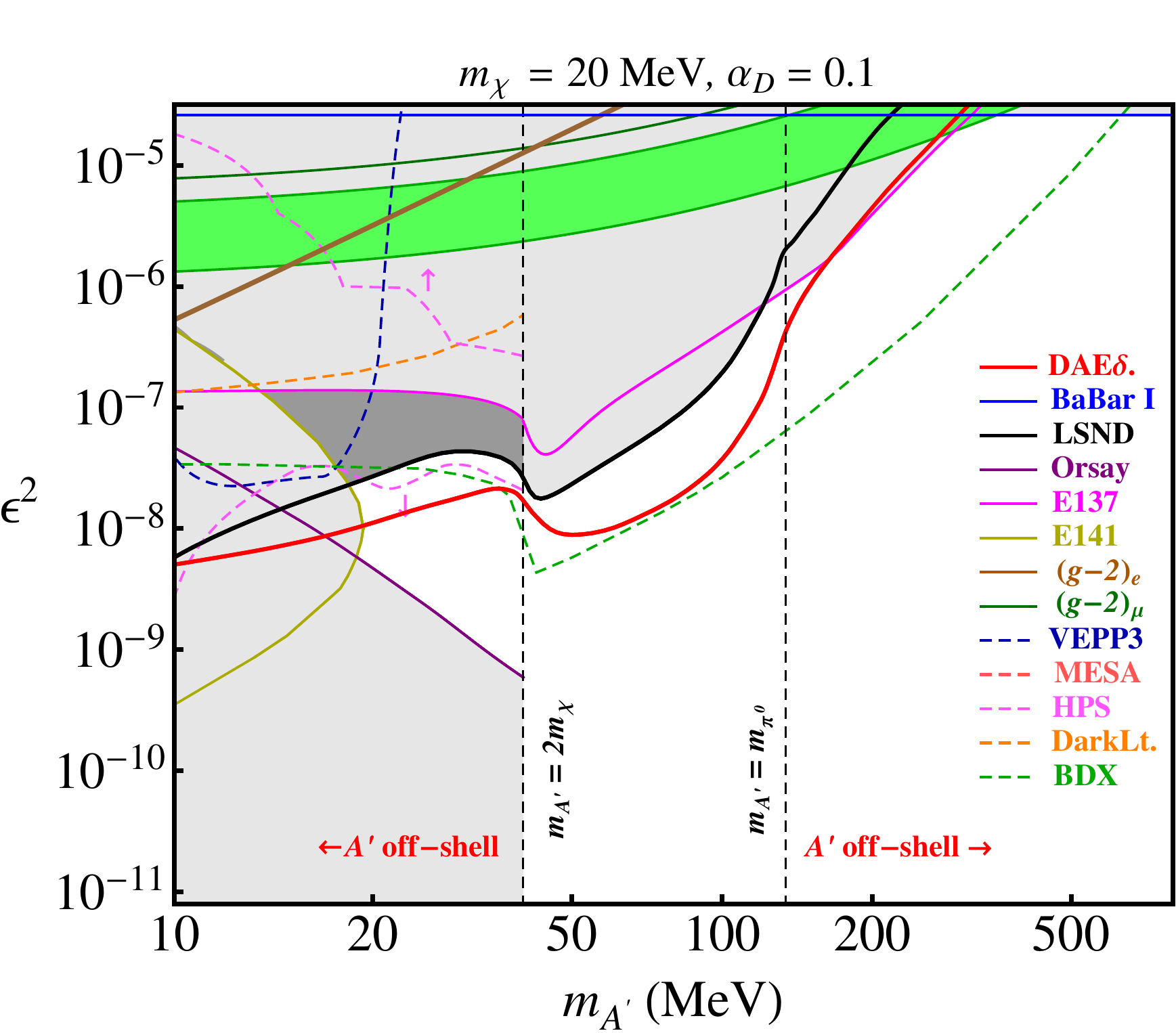}}  ~~~~~~~~
 \subfloat[]{ \label{fig:money:d}   \includegraphics[width=7.45cm]{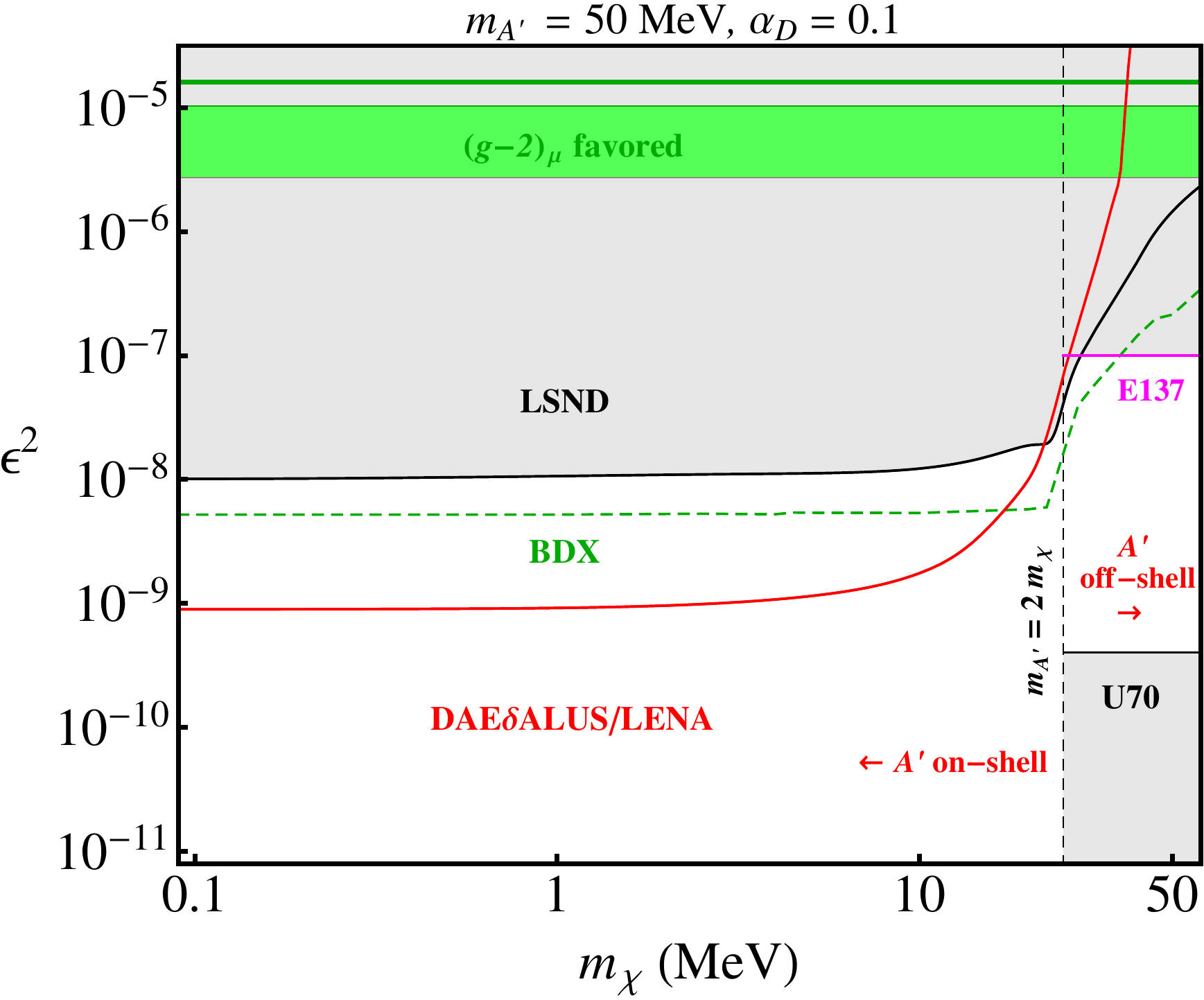}} \\
     \vspace{-0.4cm}
 \subfloat[]{ \label{fig:money:e} \includegraphics[width=7.45cm]{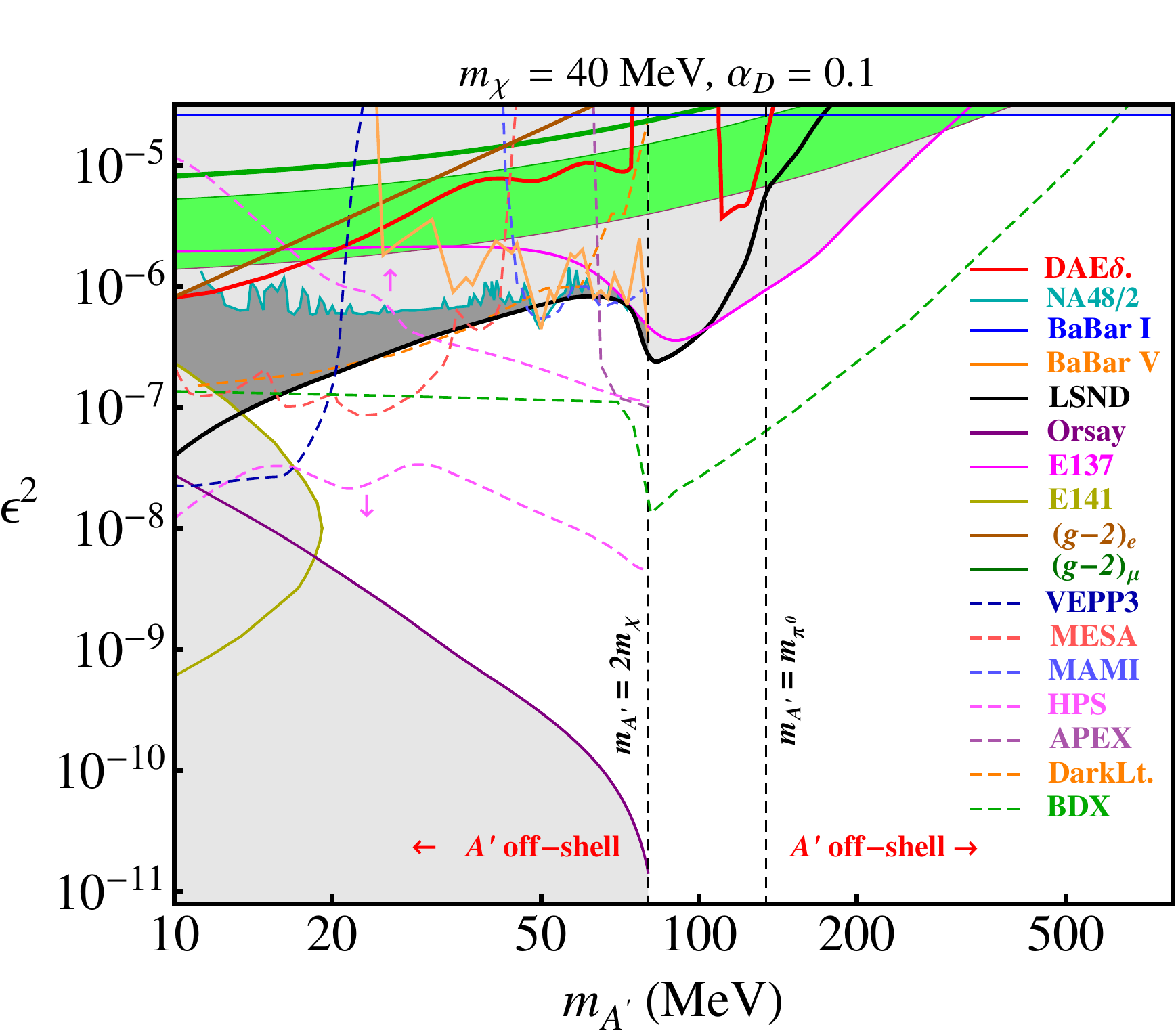}}  ~~~~~~~~
\subfloat[]{ \label{fig:money:f} \includegraphics[width=7.45cm]{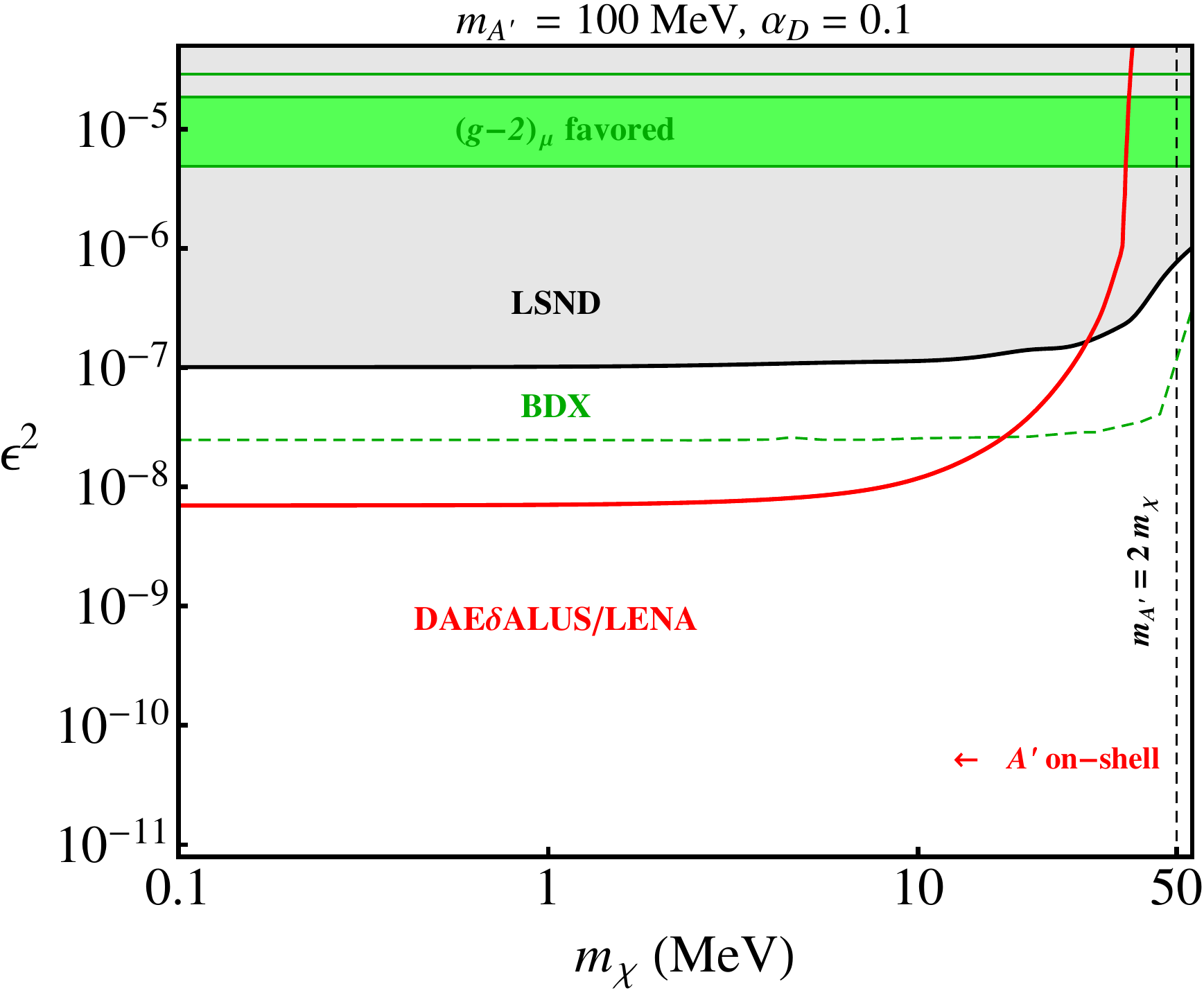}}
\vspace{-0.2cm}
     \caption{Summary of \dae/LENA $3\sigma$ sensitivity to the kinetic mixing parameter $\epsilon^2$ assuming the on-axis configuration (see \Fig{fig:LENAGeometries}c) and a full year of run-time with $7.5 \times 10^{22}$ $\pi^0$ produced. We also display updated bounds from existing LSND data in both off-shell $A'$ regimes. {\bf Left column:} \dae\ sensitivity as a  function of  $m_{\apr}$ for fixed DM mass $m_\chi = $ 1 MeV (a), 20 MeV (c), and 40 MeV (e). {\bf Right column:} \dae\ sensitivity as a  function of $m_\chi$ for fixed dark photon mass $m_{\apr} = $ 10 MeV (b), 50 MeV (d), and 100 MeV (f).  The thick green band is the region where $\apr$ could resolve the long-standing $(g-2)_\mu$ anomaly to within $\pm 2\sigma$  \cite{Pospelov:2008zw}; see \Sec{sec:Sensitivity} for information about the other projected sensitivities and constraints. Where applicable, the dashed vertical black line marks the transition between the on- and off-shell $A'$ regimes for $\pi^0 \to \gamma A^{\prime(*)} \to \gamma \chi \overline{\chi}$. In the lower off-shell regime, where we compare to visible $A' \to e^+ e^-$ searches, we emphasize that the LSND and \dae\ limits assume the existence of the off-shell process $A^{\prime*} \to \chi \overline{\chi}$.  Assuming such a $\chi$ exists, the dark gray region above the black LSND curve is excluded; this is the first demonstration that LSND can rule out a {\it visibly} decaying $A'$ by searching for DM produced via an off-shell $A^{\prime}$.
  }\label{fig:money}
\vspace{0cm}
\end{figure}
% -----------------------------------------------------------------------------------------------------------------------------------------------------------------

\begin{figure}[t!] 
 \includegraphics[width=11cm]{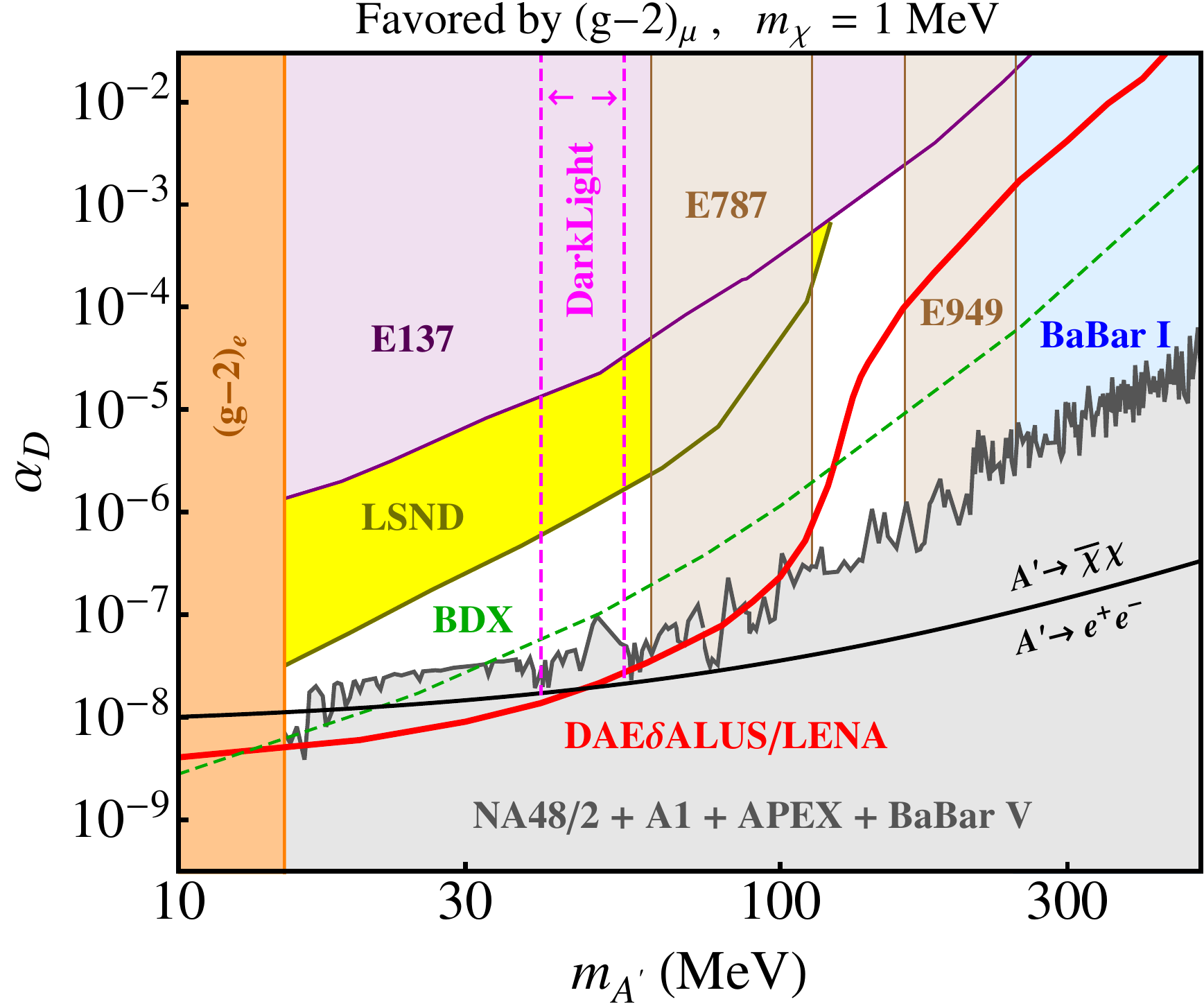} 
 \caption{Parameter space for the dark photon mass $m_{\apr}$ and dark coupling $\alpha_D$, taking $\epsilon$ to be the {\it smallest} value which resolves the $(g-2)_\mu$ anomaly for $m_\chi = 1$ MeV. The \dae/LENA curve shows $3\sigma$ sensitivity. The solid black curve is the boundary where $\Br(A'\to e^+e^-) = \Br(A'\to \bar \chi\chi) = 50 \%$. Note that for $\Br(\apr \to e^+e^-) \simeq 100 \%$ (just below the black curve) recent (preliminary) results from NA48/2 \cite{NA48} have ruled out the remaining parameter space for a visibly decaying $\apr$ that explains the discrepancy.
  }\label{fig:g2favored}
\end{figure}

To directly compare to previous studies \cite{Essig:2010xa,Kahn:2012br,Diamond:2013oda,deNiverville:2012ij,Essig:2009nc,deNiverville:2011it,Izaguirre:2014dua,Izaguirre:2013uxa,Freytsis:2009bh,Batell:2009di,Morrissey:2014yma}, we will focus on vector portal models of the dark sector \cite{Holdom:1985ag, Okun:1982xi, Galison:1983pa}.  Here, a massive dark photon $A'$ from a new U(1)$_D$ kinetically mixes with the standard model hypercharge:\footnote{The $A'$ can acquire mass either through a St\"uckelberg field or a dark Higgs.}
\be
\mathcal{L} \supset  \frac{\epsilon_Y}{2} F'_{\mu \nu}B^{\mu \nu} + \frac{m_{A^\prime}^2}{2} A^\prime_\mu A^{\prime \mu}+ \bar \chi (i \displaystyle{\not}{D} - m_\chi ) \chi , 
\ee
and couples to a DM particle $\chi$, which carries unit charge under the U(1)$_D$. The DM can be either a scalar or a Dirac fermion; we focus in the text on the case of fermionic DM, leaving a discussion of scalar DM to the appendices. Here, $D_\mu \equiv \partial_\mu + i g_D A_{\mu}^{\prime} $, where $g_D$ is the 
dark coupling constant. 
After electroweak symmetry breaking and diagonalizing the kinetic terms, the $A'$ inherits a universal coupling to electromagnetic currents with strength $\epsilon e$, where $\epsilon \equiv \epsilon_Y \cos \theta_W$. This model has four free parameters,
\be
\{ m_{A'}, \epsilon, m_\chi, \alpha_D \},
\ee
namely the $A'$ mass $m_{A'}$, the kinetic mixing parameter $\epsilon$, the DM mass $m_{\chi}$, and the dark fine-structure constant $\alpha_D \equiv g_D^2/4\pi$.  Many dark photon studies have explored the $\{m_{A'}, \epsilon\}$ portion of parameter space, but $m_\chi$ is an essential third dimension that introduces qualitatively different phenomenology. We focus primarily on the region of parameter space $\alpha_D \gg \epsilon^2 \alphaEM$ where the $A'$ primarily decays into DM when kinematically allowed, rather than into visible-sector particles, though we do look at a wider range of $\alpha_D$ values in \Fig{fig:g2favored}.\footnote{Changing $\alpha_D$ results in a simple linear scaling of the sensitivity when the DM is produced via an on-shell $A'$, and a quadratic scaling when the DM is produced via an off-shell $A'$. We discuss scaling with $\alpha_D$ in \Sec{sec:Sensitivity}.}

% -----------------------------------------------------------------------------------------------------------------------------------------------------------------
%						Model Independent g-2 Coverage 
% -----------------------------------------------------------------------------------------------------------------------------------------------------------------

% -----------------------------------------------------------------------------------------------------------------------------------------------------------------

Due to its universal coupling to electromagnetism, the $A'$ can replace a photon in any kinematically-allowed process, with an accompanying factor of $\epsilon$, such that the event rate for any tree-level process coupling the visible sector to the dark sector is proportional to $\epsilon^2$. Thus, as shown in \Fig{fig:production}, DM can be produced and detected via
\begin{align}
\pi^0 & \to \gamma A^{\prime(*)} \to \gamma \chi \overline{\chi},\\
\chi e^- & \to \chi e^-,
\end{align}
where the $A'$ can either be on- or off-shell in the production process, and the scattering process proceeds through a $t$-channel $A'$.\footnote{Since $\chi$ and $\overline{\chi}$ are indistinguishable in the detector, we only write $\chi$ for simplicity.} The main detection backgrounds come from neutrinos, either elastic scattering off electrons or charged-current quasi-elastic (CCQE) scattering off nucleons, but because the spectra of neutrinos produced from decays at rest have sharp kinematic cutoffs, much of the neutrino background can be mitigated by a simple cut on the electron recoil energy in the detector.

While our benchmark dark sector is a viable, renormalizable theory of DM in its own right, it is also useful to regard this scenario as a simplified model for an entire class of theories in which sub-GeV particles mediate interactions between dark and visible matter.  Indeed, there is a vast literature which invokes light, weakly-coupled particles to resolve anomalies in direct and indirect detection experiments, build models that relate dark and baryonic energy densities, resolve puzzles in simulations of cosmological structure formation, introduce new relativistic degrees of freedom during big bang nucleosynthesis, and resolve the proton charge-radius anomaly and other low-energy standard model anomalies \cite{Okun:1982xi,Galison:1983pa,Holdom:1985ag,Boehm:2003hm,Boehm:2003bt,Finkbeiner:2007kk,Huh:2007zw,Pospelov:2007mp,Pospelov:2007mp,Kahn:2007ru,Hooper:2008im,Feng:2008ya,Feng:2008mu,ArkaniHamed:2008qn,Nomura:2008ru,Pospelov:2008zw,Alves:2009nf,Cirelli:2009uv,Morrissey:2009ur,Mardon:2009gw,Chang:2010yk,Barger:2010aj,TuckerSmith:2010ra,Falkowski:2011xh,Kaplan:2011yj,Essig:2011nj,Andreas:2011in,Morris:2011dj,Cline:2012is,An:2012va,Graham:2012su,Hooper:2012cw,CyrRacine:2012fz,Brust:2013ova,Kaplinghat:2013xca,Foot:2014mia,Shuve:2014doa,Krnjaic:2014xza,Izaguirre:2014cza,Detmold:2014qqa}. That said, it has been observed that certain realizations of light ($\lesssim$ GeV) DM face strong constraints from out-of-equilibrium annihilation to charged leptons during CMB freeze-out \cite{Slatyer:2009yq,Galli:2009zc,Cirelli:2009bb,Galli:2011rz,Hutsi:2011vx}. However, these bounds are model dependent and can be evaded if DM is asymmetric, scatters inelastically with the visible sector \cite{Izaguirre:2014dua}, has a velocity-suppressed annihilation cross section \cite{Lin:2011gj}, or if the annihilating particles are a subdominant fraction of the DM abundance, none of which affect the projections for a fixed target search.\footnote{A thorough analysis of model-dependent cosmological constraints is beyond the scope of this work, but see \Ref{Izaguirre:2013uxa} for a more in-depth discussion of these issues. We simply note here that in the region of parameter space we consider, $\alpha_D$ is typically large enough to make the relic density of $\chi$ a subdominant fraction of the observed total DM abundance.}  We therefore consider the kinetically-mixed dark photon as a simplified model of a portal to the dark sector for which the experimental constraints and future projections can be adapted to study a plethora of other, more elaborate scenarios.  

The rest of this paper is organized as follows. In \Sec{sec:Production}, we describe the mechanism of DM production at the \dae\ source, for both on- and off-shell mediators. We describe the mechanism and signals of DM scattering at the LENA detector in \Sec{sec:Scattering}, and we survey the backgrounds to such a search in \Secs{sec:BeamOffBackgrounds}{sec:BeamOnBackgrounds}. In \Sec{sec:Sensitivity}, we discuss the sensitivity of \dae/LENA to DM production in various regions of parameter space, and compare with re-evaluated bounds from LSND and limits from searches for $A' \to e^+ e^-$. We conclude in \Sec{sec:Conclusion}. Details of the various production and scattering calculations can be found in the appendices.

\section{Dark Matter Production at \dae}
\label{sec:Production}

As mentioned above, production of dark photons $A'$ can be achieved by replacing a photon with an $A'$ in any kinematically-allowed process. At the 800 \MeV \ proton kinetic energies of the \dae\ beam, photons come primarily from $\pi^0$ decays, where the pions are produced mostly from $\Delta$ resonances:
\begin{equation}
\Delta^+ \to p + \pi^0, \hspace{3mm} \Delta^0 \to n + \pi^0.
\end{equation}
$A'$s can also be produced directly from radiative $\Delta$ decays, $\Delta \to N + A'$, where $N$ is a proton or neutron. The branching ratio for $\Delta \to N + \gamma$ is approximately $0.5\%$, and so is subdominant to $A'$ production from pion decays, except in the range $m_{\pizero} < m_{A'} < m_\Delta - m_N$ where the $A'$ is on-shell from $\Delta$ decay but off-shell from $\pizero$ decay. Lacking a reliable way to simulate $\Delta$ production and decay, we neglect this signal mode in our analysis, though we estimate that it may improve signal yield by as much as a factor of 2 over the range $m_{A'} \in [135, 292] \ \MeV$.\footnote{We thank Rouven Essig for pointing out the importance of on-shell $A'$ production from $\Delta$ decays.} Other sources of photons are expected to be negligible for our sensitivity estimates: $\rho$ and $\eta$ mesons are kinematically inaccessible, and bremsstrahlung photons produced in the hadronic shower are suppressed by $\alphaEM$, $m_p$, and phase space factors, making them subdominant to photons from $\Delta$ decays. Consequently, we will focus on DM production through $\pi^0 \to \gamma A^{\prime(*)}  \to \gamma \chi \overline{\chi}$, where the $A'$ can be either on- or off-shell depending on the masses of the DM and the $A'$.

We simulated DM production by obtaining a list of $\pi^0$ events from GEANT 4.9.3 \cite{Agostinelli:2002hh} with a simplified model of the \dae\ target geometry, and generated the DM kinematics by decaying the pions as predicted by the dark photon model; details are given below and in \App{app:Production}.\footnote{We used the ``QGSP\_BIC'' physics list in GEANT4.} Previous studies \cite{Batell:2009di, deNiverville:2011it} have assumed that the $\pi^0$ energy spectrum from proton-carbon collisions is similar to the $\pi^+$ spectrum, and used fits to $\pi^+$ data \cite{Burman:1989ds} to model the $\pi^0$ production. We find reasonable agreement with this assumption based on the GEANT simulation, though the spectra of $\pi^+$ versus $\pi^0$ differ considerably at high energies. Similarly, in previous studies, the total $\pi^+$ production rate was estimated by working backwards from the observed neutrino flux within the detector acceptance, and assuming that all neutrinos came from $\pi^+$ decays at rest; the $\pi^0$ total rate was assumed to be equal to the $\pi^+$ rate up to a factor of 2 uncertainty \cite{deNiverville:2011it}. In our approach, the same GEANT simulation can simulate both $\pi^0$ and $\pi^+$ production, allowing an estimate of the $\pi^0$ rate which does not rely on such assumptions about the $\pi^+$ rate.

If $2m_{\chi} < m_{A'} < m_{\pizero}$, the $A'$ can be produced on-shell and decay to DM. The narrow width approximation \cite{SchwartzQFT} can be used to obtain a simple expression for the branching ratio, 
\begin{equation}
\label{BRDM}
\Br(\pizero \to \gamma \chi \overline{\chi}) = \Br(\pizero \to \gamma \gamma) \times 2\epsilon^2 \left (1 - \frac{m_{A'}^2}{m_{\pizero}^2} \right)^3 \times \Br(A' \to \chi \overline{\chi}) \hspace{3mm} \textrm{(on-shell)}.
\end{equation}
In the region of parameter space where $\alpha_D \gg \epsilon^2 \alphaEM$, $\Br(A' \to \chi \overline{\chi}) \approx 1$. Then $\Br(\pizero \to \gamma \chi \overline{\chi})$ is independent of $m_\chi$ and $\alpha_D$ and depends only on the $A'$ mass and the kinetic mixing parameter $\epsilon$. Since the kinematics of two-body decays are fixed by energy-momentum conservation, the double-differential angular and energy distribution $d^2 N_\chi/ (d\Omega \, dE_\chi)$ (summed over the DM polarizations and the unobserved photon polarizations) of the DM is also independent of $m_\chi$, and is inherited directly from the analogous distribution of the $A'$s, which is, in turn, inherited from the parent pions. However, we caution that the narrow-width approximation breaks down if $m_{A'}$ is sufficiently close to $m_{\pizero}$ from below \cite{Berdine:2007uv,Kauer:2007nt,Uhlemann:2008pm}. In particular, there is no sharp kinematic threshold at $m_{\pizero}$.

If $m_{A'} < 2m_\chi$ or $m^2_{A'} \gtrsim m^2_{\pizero} - 2\Gamma_{A'} m_{A'}$, the narrow-width approximation is not applicable, and DM is produced through a three-body decay.\footnote{This illustrates a subtlety of the narrow-width approximation. Although the $A'$ \emph{can} go on-shell for $m_{A'} < m_{\pizero}$, the phase space suppression means that the phase space integral in \Eq{BRDMOffshell} is actually dominated by the off-shell region of the amplitude, giving a smooth behavior through the $\pizero$ threshold. The effect of near-degeneracies on the efficacy of the narrow width approximation in resonant three-body decays has been previously noted in \Ref{Uhlemann:2008pm}, where it is shown that phase-space factors distort the shape of the Breit-Wigner and lead to errors parametrically greater than $\Gamma/M$.}  Details of our treatment of the narrow width approximation are given in \Apps{app:OnShell}{app:Threshold}.
The expression for the branching ratio involves a phase-space integral which cannot be computed analytically,
\begin{equation}
\label{BRDMOffshell}
\Br(\pizero \to \gamma \chi \overline{\chi}) = \frac{1}{\Gamma_{\pizero}} \times \frac{\epsilon^2 \alpha_D}{2 m_{\pizero}} \int d \Phi_{\pizero \to \gamma A'} \, d\Phi_{A' \to \chi \overline{\chi}} \,\frac{ds}{2\pi} \langle |\mathcal{\hat{A}}_{\pizero \to \gamma \chi \overline{\chi} }|^2 \rangle \hspace{3mm} \textrm{(off-shell)},
\end{equation}
where $s$ is the mass-squared of the virtual $A'$, $\Gamma_{\pizero} = 7.74 \ \eV$ is the total $\pizero$ width, and $\mathcal{\hat{A}}_{\pizero \to \gamma \chi \overline{\chi}}$ is the three-body decay amplitude normalized to $\epsilon = \alpha_D = 1$. This normalization was chosen to make the dependence of the branching ratio on $\epsilon$ and $\alpha_D$ explicit.  In contrast to the on-shell case, the branching ratio now depends on both the dark fine structure constant $\alpha_D$ and the DM mass $m_\chi$. Full expressions for the three-body amplitudes for fermionic and scalar $\chi$, as well as the $A'$ width, are given in \App{app:Production}. The double-differential distribution $d^2N_\chi/(d\Omega \, dE_\chi)$ can be obtained in a straightforward manner from \Eq{BRDMOffshell} by only performing the first phase space integral, which gives the distribution in the $\pi^0$ rest frame, then boosting according to the $\pi^0$ lab-frame distribution.

Putting these pieces together, the total number of DM particles produced at \dae\ is
\begin{equation}
\label{eq:DMFlux}
N_\chi = 2N_{\pizero} \, \Br(\pizero \to \gamma \chi \overline{\chi}),
\end{equation}
where our GEANT simulation yields  $N_{\pi^0} = 7.5 \times 10^{22}$ $\pizero$/yr, and $\Br(\pizero \to \gamma \chi \overline{\chi})$ is given by \Eq{BRDM} for on-shell production and \Eq{BRDMOffshell} for off-shell production. The maximum energy of DM produced at \dae\ as a function of its mass $m_\chi$ is
\begin{equation}
\label{eq:DMMaxE}
E_{\chi}^{\rm max} = \frac{1}{2}\gamma_{\rm max} m_{\pizero} \left (1 + \beta_{\rm max}\sqrt{1 - \frac{4m_\chi^2}{m_{\pizero}^2}}\right),
\end{equation}
where $(\gamma_{\rm max}, \beta_{\rm max}) \simeq (5,0.98)$ are the maximum boost and velocity respectively for $\pizero$s produced at \dae.

\section{Dark Matter Scattering at LENA}
\label{sec:Scattering}

% -----------------------------------------------------------------------------------------------------------------------------------------------------------------
% 						Geometric comparison
% -----------------------------------------------------------------------------------------------------------------------------------------------------------------

\begin{figure}[t!] 
 \hspace{-0.5cm}
 \includegraphics[width=10cm]{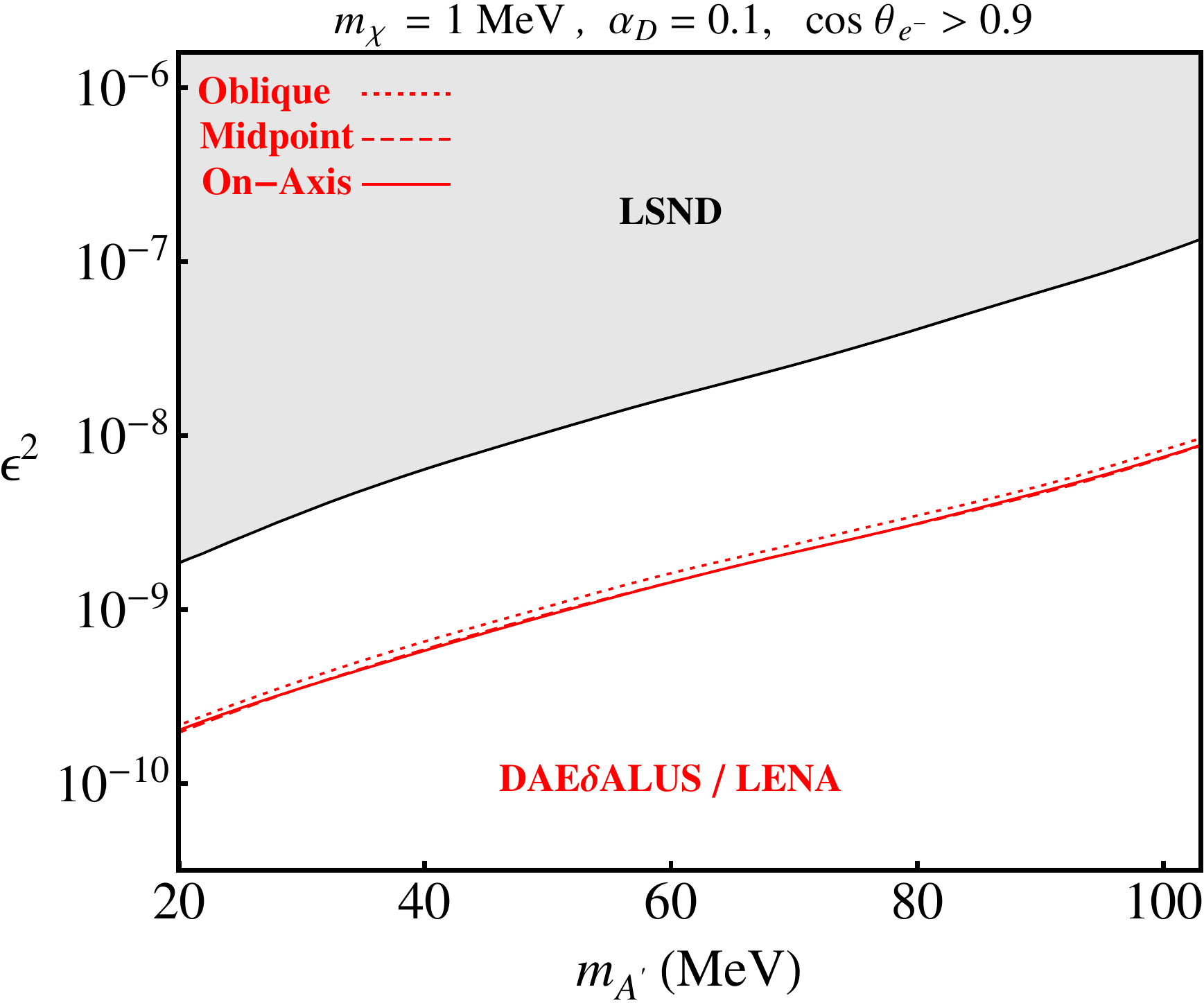}
  \caption{Sensitivity contours at \dae/LENA showing the effect of changing experimental geometries. All curves assume a 3$\sigma$ signal-to-background sensitivity, see \Secs{sec:BeamOnBackgrounds}{sec:BeamOffBackgrounds}. Existing limits from the multi-year data set at LSND \cite{deNiverville:2011it} are shown for comparison. The signal contours are computed by integrating the electron recoil profile over the interval
that maximizes $S/\delta B$ for each value of $m_{A'}$. 
}\label{fig:ChangeGeo}
\vspace{0cm}
\end{figure}
% -----------------------------------------------------------------------------------------------------------------------------------------------------------------

The LENA detector \cite{Wurm:2011zn} is a proposed cylindrical scintillator detector with a target volume of radius 13 m and height 100 m; we assume the target volume is filled with linear-alkyl-benzene (C$_{18}$H$_{30}$), giving a fiducial mass of 45.8 kiloton, though other choices of scintillator are under consideration. Dark sector particles produced at the \dae\ target can travel unimpeded through the surrounding material to scatter in the LENA detector.  For low mass
mediators, the dominant channel is coherent scattering off detector nuclei, which enjoys an $A^2$ enhancement since small momentum transfers are unable to resolve
nuclear substructure. However, this channel suffers from a severe form-factor suppression for momentum transfers in excess of our electron recoil cuts which are necessary to discriminate the signal from the beam-on neutrino backgrounds. DM particles can also scatter off atomic electrons in the detector, and it is this $\chi e^- \to \chi e^-$ channel which we will focus on, though the discussion below can be adapted to a generic detector target.\footnote{If there are mass splittings in the dark sector and the $A^\prime$ coupling is off-diagonal between mass eigenstates, scattering inside the detector will be inelastic and may feature striking de-excitation signals that are not easily mimicked by neutrino or cosmic backgrounds \cite{Izaguirre:2014dua}. Although this scenario is beyond the scope of this paper, we note that the experimental setups discussed in this work should have promising discovery potential for these signals as well, and in \App{app:Scattering} we derive cross sections appropriate to this more general case.}

The total scattering yield for the electron channel is
\be
\label{eq:yield}
N_{\rm sig} = n_e \int_{ E_e^{\rm low}(m_\chi)}^{E_e^{\rm high}(m_\chi)}  dE_{e} \int_{E_{\chi}^{ \rm min}(E_{e})  }   \!\! dE_{\chi}  \, \int_{\rm LENA}  d\Omega \, \ell(\Omega)
\frac{d^2 N_{\chi}}{ d\Omega \, dE_{\chi}}   \frac{d\sigma}{dE_{e}},
\ee
where $n_e = 3.0 \times 10^{23}/{\rm cm}^3$ is the number density of target electrons, $\ell(\Omega)$ is the DM path length through LENA, $d\sigma/dE_e$ is the recoil electron energy distribution, and the angular integral is taken over the region covered by the LENA detector for the chosen geometry. $E_e^{\rm low}(m_\chi)$ and $E_e^{\rm high}(m_\chi)$ are electron recoil energy cuts which are chosen for each $m_\chi$ to optimize signal-to-background sensitivity for that mass point; we discuss these cuts further in \Sec{sec:BeamOnBackgrounds}. In principle, we should also include a factor accounting for any muon veto dead time or reconstruction efficiencies, but we neglect these here. The minimum incoming energy for $\chi$ to induce an electron recoil of energy $E_{e}$ is
\be
\label{eq:minimumChiEnergy}
 E_{\chi}^{\rm min}(E_{e}) =  \frac{T_e}{2} \left[
1 +  \sqrt{     \left(1 + \frac{2 m_e}{T_e} \right)  \left(1 + \frac{2 m_\chi^2}{ m_e T_e} \right)     
}                    \,   \right],  \qquad T_e \equiv E_e - m_e,
\ee
where $m_e$ is the electron mass and $T_e$ is the electron kinetic energy. Another useful expression is the maximum possible recoil electron energy for a given DM mass,
\be
\label{eq:maxEEnergy}
E_{e}^{\rm max}(m_\chi) = m_e + \frac{2({E_{\chi}^{\rm max}})^2 - 2m_\chi^2}{2E_{\chi}^{\rm max}\,m_e + m_\chi^2 + m_e^2},
\ee
where $E_{\chi}^{\rm max}$ is given in \Eq{eq:DMMaxE}. In \App{app:Scattering}, we present the details of our numerical signal rate computation, including cross sections for scalar and fermion DM particles scattering off a generic target.

In terms of geometry, we consider three possible locations for \dae\ relative to LENA, shown in \Fig{fig:LENAGeometries}:
\begin{itemize}
\item \textit{midpoint}---pointed horizontally at the vertical midpoint of the detector, 16 m away from the cylindrical face;
\item \textit{oblique}---pointed horizontally near the upper corner of the detector, at a lateral distance 16 m and height 5 m;
\item \textit{on-axis}---pointed downwards into the endcap of the detector, 16 m above the top face.
\end{itemize}
The LENA design is self-shielding and includes a 2 m buffer and 2 m muon veto between the outer face and the target volume, so the effective source-detector distance in all three cases is at least 20 m. The signal yield for a 1 \MeV \ DM particle for the three proposed geometries is shown in \Fig{fig:ChangeGeo}.  The choice of geometry only affects the sensitivity in $\epsilon^2$ by a factor of order 10\%. The midpoint and on-axis geometries are essentially identical, and provide superior sensitivity compared to the oblique geometry for the entire range of $A'$ masses; the effective detector length and solid angle acceptance are larger for these geometries, and because the signal and background angular distributions are so similar after energy cuts are imposed (see \Fig{fig:SigvsBGDistributions:a} and the discussion below), no additional signal/background separation is achieved in the oblique configuration. For simplicity, we will focus on the on-axis configuration because it preserves cylindrical symmetry.

In terms of electron energy cuts, we consider three benchmark cuts on $E_{e}$ based on avoiding various beam-on background thresholds:
\begin{itemize}
\item $E_e^{\rm low} = 106~\MeV$---Above the low-energy muon capture and stopped pion and muon backgrounds;
\item $E_e^{\rm low} = 147~\MeV$---Above the energy threshold for muon production from beam-on sources;
\item $E_e^{\rm low} = 250~\MeV$---Above the dominant decay-in-flight neutrino-electron scattering background.
\end{itemize}
Roughly speaking, the 106 MeV cut is optimal for heavy DM, the 147 MeV cut is optimal for medium-mass DM, and the 250 MeV cut is optimal for light DM. This can be seen from \Eq{eq:maxEEnergy}: for example, $m_\chi = 42 \ \MeV$ implies $E_e^{\rm max} = 146 \ \MeV$, so the lowest of the energy thresholds (with all its additional backgrounds) is necessary to retain any signal acceptance at all. We give more details justifying these cuts in \Sec{sec:BeamOnBackgrounds} below, and discuss how to optimize them based on the various background spectra.

\section{Beam-off Backgrounds}
\label{sec:BeamOffBackgrounds}

%----------------------------------------
%Beam-off table
%----------------------------------------
\begin{table*}[tdp]
\begin{center}\begin{tabular}{|c||c|c|c|c|c|c|}
\hline \hline
\textbf{Source} & \textbf{Neutrino} &  \textbf{Reaction Type}  & \textbf{106--147 MeV} & \textbf{147--250 MeV} & \textbf{250--400 MeV} & \textbf{Tag}  \\
\hline
\hline
\multirow{8}{*}{Atmospheric}
& \multirow{2}{*}{$\nu_\mu$}
& elastic &  \hspace{-0.3cm}$<1$ &  \hspace{-0.3cm}$<1$ &  \hspace{-0.3cm}$<1$ & -- \\
\cline{3-7}
& & CCQE & 6 & 13 & 12 & Michel \\
\cline{2-7}
& \multirow{2}{*}{$\nu_e$}
& elastic &  \hspace{-0.3cm}$<1$ &  \hspace{-0.3cm}$<1$ &  \hspace{-0.3cm}$<1$ & -- \\
\cline{3-7}
& & CCQE & 3 & 9 & 9 & -- \\
\cline{2-7}
& \multirow{2}{*}{$\overline{\nu}_\mu$}
& elastic & \hspace{-0.3cm}$<1$ & \hspace{-0.3cm}$<1$ & \hspace{-0.3cm}$<1$ & -- \\
\cline{3-7}
& & CCQE & 2 & 4 & 4 & Michel \\
\cline{2-7}
& \multirow{2}{*}{$\overline{\nu}_e$}
& elastic &  \hspace{-0.3cm}$<1$ &  \hspace{-0.3cm}$<1$ &  \hspace{-0.3cm}$<1$ & -- \\
\cline{3-7}
& & CCQE & 1 & 2 & 2 & neutron\\
\hline
\hline
\end{tabular}
\caption{One-year rates for all beam-off backgrounds resulting in an outgoing lepton $\ell = e, \mu$ with kinetic energy $T_\ell > 106 \ \MeV$ in the final state. ``Elastic'' refers to elastic neutrino-electron scattering, and ``CCQE'' refers to charged-current quasi-elastic neutrino-nucleon scattering. A cut $\cos \theta_\ell > 0.9$ has been imposed on all outgoing charged leptons.}
\label{tab:AllBeamOffBackgrounds}
\end{center}
\end{table*}

The signal process $\chi e^- \to \chi e^-$ faces backgrounds from any process which results in an energetic lepton in the final state. There are two main sources of backgrounds, beam-off and beam-on. The principal advantage of using an underground detector such as LENA is the reduction in beam-off backgrounds from sources other than neutrinos. The target depth of LENA is approximately 4000 m.w.e.\ with a cosmic muon flux of $\simeq 1 \times 10^{-4} {\rm m}^{-2} {\rm s}^{-1}$.  Therefore, external backgrounds related to untagged cosmic muons interacting in the rock surrounding the detector are expected to be negligible in our energy range of interest, $E>106~$MeV.  Consequently, we focus only on backgrounds involving neutrinos.
Elastic neutrino-electron scattering from atmospheric neutrinos of any flavor, 
\be
\nu e^- \to \nu e^-,
\ee
poses an irreducible beam-off background since it has the same final state as the signal process. However, there is an additional type of background from charged-current quasi-elastic (CCQE) scattering of neutrinos, 
\be
\nu_\ell \, n \to \ell^- \, p, \qquad \overline{\nu}_\ell \, p \to \ell^+ \, n.
\ee
Despite the fact that this event has a completely different final state from the signal process (with for example hadronic activity in addition to the lepton), for $\nu_e$ this process is an \emph{irreducible} background at LENA because the energy from the vertex activity cannot be separated from the energy of the produced electron.\footnote{In principle, events with delayed vertex activity such as $\nu_e \ {}^{12}{\rm C}  \to e^-  \ {}^{12}N_{\rm gs}, {}^{12}N_{\rm gs}\rightarrow {}^{12}{\rm C}~\beta^+$ can be tagged, but we do not consider event-by-event rejection of this class of events here.}  For all other neutrino flavors, this process is at least partially reducible, by detecting the Michel electron from the muon decay for $\ell = \mu^\pm$, and by tagging the neutron for $\ell = e^+$ when the CCQE reaction takes place on hydrogen. However, since the duty cycle of the \dae\ cyclotron is only 25\%, all of these backgrounds can be measured directly during beam-off time and then scaled to the beam-on time with a systematic uncertainty of $\sqrt{3B}/3$.  This is combined in quadrature with the statistical uncertainty $\sqrt{B}$ on the background during beam-on time, giving a total background uncertainty which scales as $\delta B = \sqrt{4B/3}$.

The spectrum of atmospheric neutrinos extends to very high energies, so to reduce the rate of high-energy neutrino scattering feeding down into lower electron recoil energies, we will impose a maximum recoil energy $E_e^{\rm max}$ for the recoil electron depending on the DM mass (see below). Furthermore, the resultant lepton is produced nearly isotropically, while high-energy electrons from DM scattering are principally scattered in the direction of the initial proton beam, as shown in \Fig{fig:SigvsBGDistributions:a}. By requiring the outgoing lepton to be within $25^\circ$ of the beamline ($\cos \theta_\ell > 0.9$) and exploiting the directional detection capabilities of LENA, we can further reduce the beam-off background while keeping $\approx 99\%$ of the signal over most of the kinematically-allowed parameter space.\footnote{LENA is able to resolve paths of outgoing electrons with energies above 250 MeV and muons with kinetic energies above 100 MeV to an accuracy of a few degrees \cite{Wurm:2011zn}. Extending this cut for electrons down to energies of 106 MeV is perhaps optimistic at LENA, but may be possible with a future detector paired with the \dae\ source.} The rates for these processes in three benchmark energy ranges of interest are given in \Tab{tab:AllBeamOffBackgrounds}; more details of our beam-off estimates are given in \App{app:NeutrinoBGBeamOff}. We note that with these cuts, all the beam-off backgrounds are subdominant to the beam-on backgrounds, which we discuss below. 

\section{Beam-on Backgrounds}
\label{sec:BeamOnBackgrounds}

% -----------------------------------------------------------------------------------------------------------------------------------------------------------------
%   mchi = 1, 20 MeV, mA = 50 MeV on-shell angular and recoil distributions
% -----------------------------------------------------------------------------------------------------------------------------------------------------------------

\begin{figure}[t!] 
 \hspace{-0.3cm}
\subfloat[]{\label{fig:SigvsBGDistributions:a} \includegraphics[width=9cm]{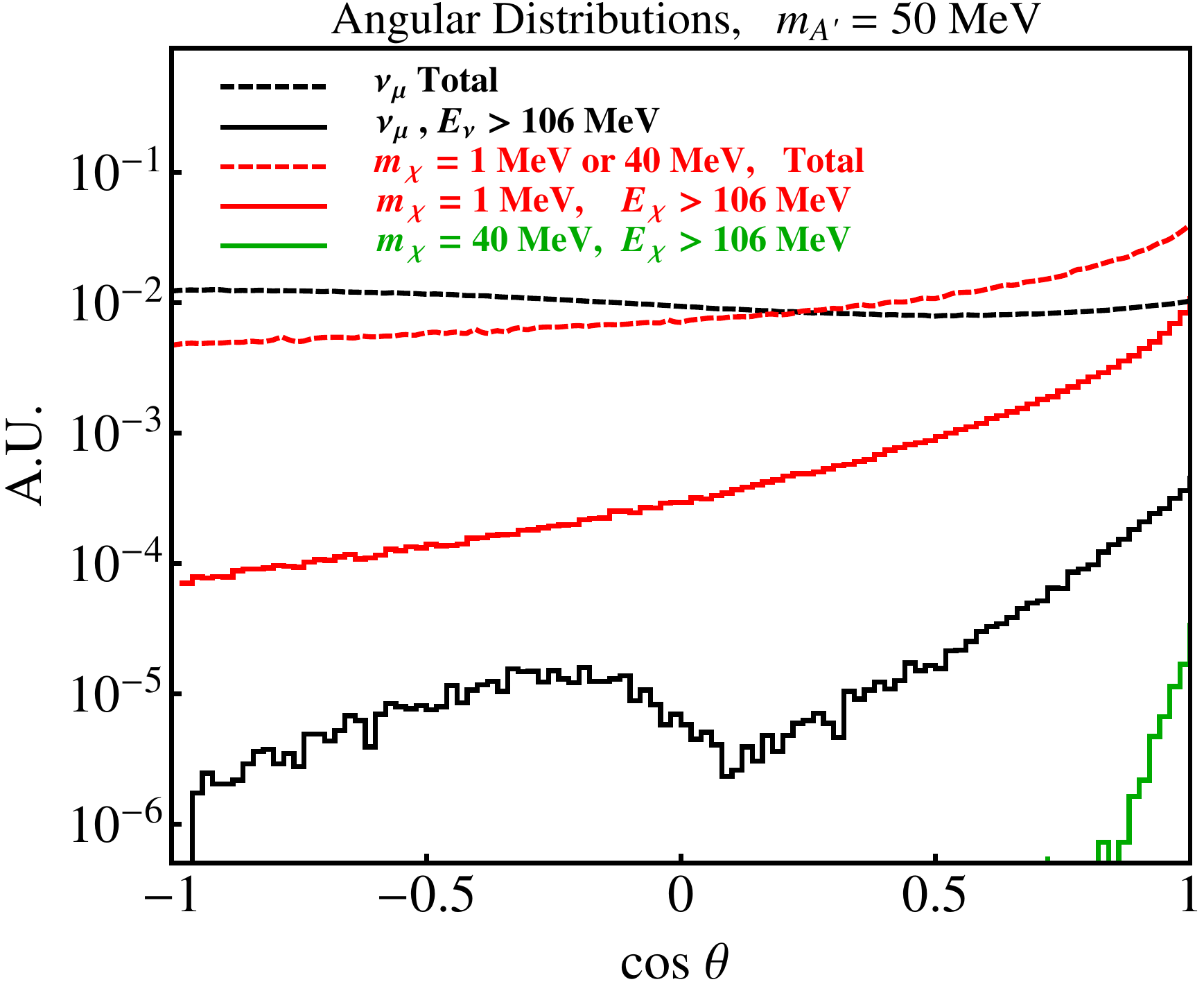}} ~ 
\subfloat[]{\label{fig:SigvsBGDistributions:b}  \includegraphics[width=8.8cm]{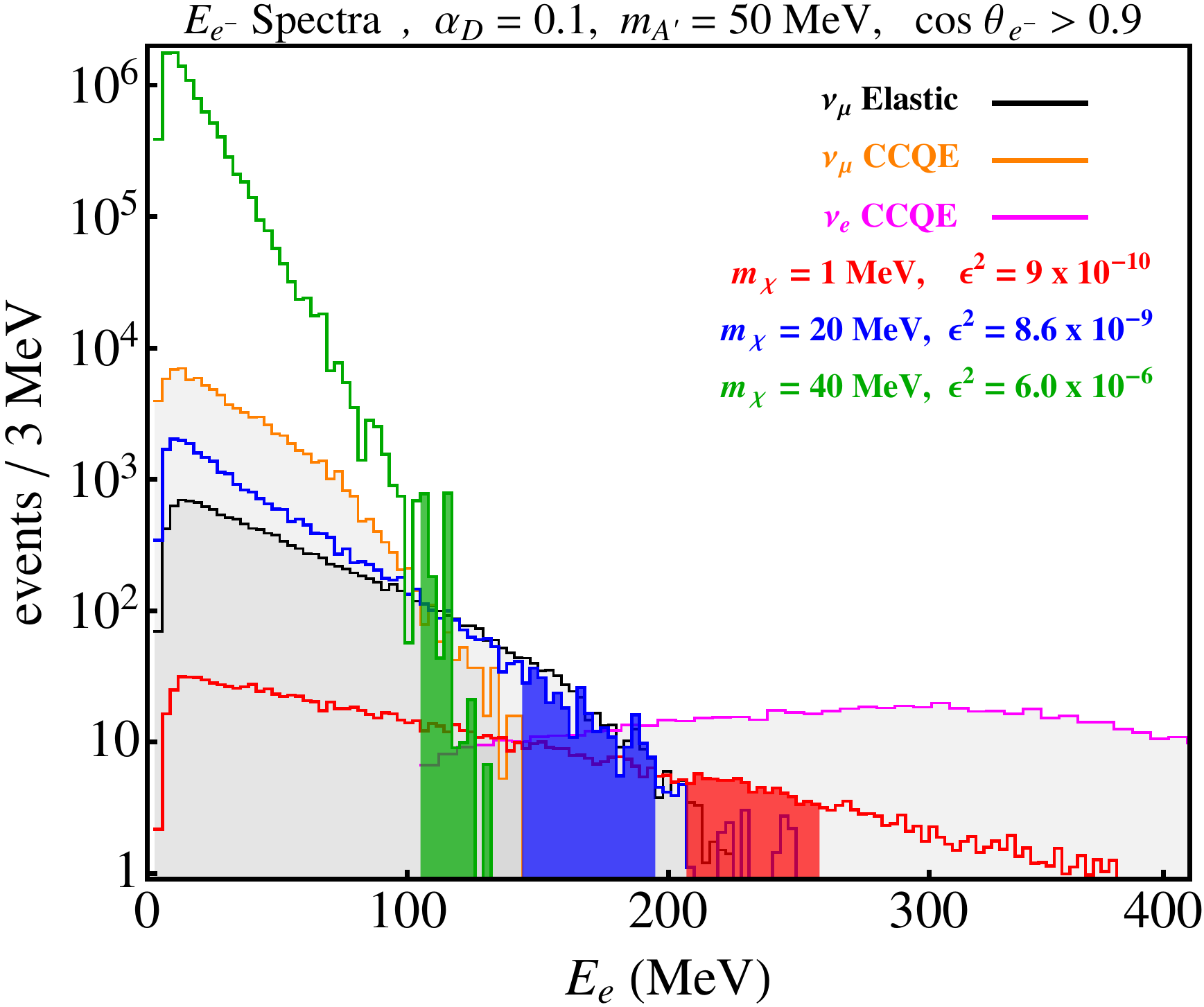}}
  \caption{{\bf Left (a):} Angular distributions for DM production and beam-on neutrinos produced at the \dae\ source. The neutrino distribution is roughly isotropic while the signal is strongly peaked in the forward direction ($\cos \theta \simeq 1$).  The slight excess of neutrino production in the backward direction is an artifact of the simplified target geometry used in the simulation; see text for details.  Above 106 MeV both the DM and neutrino distributions are strongly peaked in the forward direction; the relative normalizations of the curves with and without the cut show the reduction in signal and background due to this cut alone, though the actual signal is also determined by the geometric acceptance of LENA. For different DM masses, the normalization of the DM distribution changes, but not its shape.  
Although LENA cannot resolve electron-recoil angles for which $\cos \theta > 0.9$, imposing a stronger angular cut of $\cos\theta > 0.95$ would preserve an order-one fraction of signal events and dramatically reduce both beam-off and beam-on backgrounds discussed in \Secs{sec:BeamOffBackgrounds}{sec:BeamOnBackgrounds}. 
To be conservative, we assume $\cos\theta_\ell > 0.9$ for all of our sensitivity projections, but this is a potential avenue for improving new-physics searches in the electron
scattering channel.
{\bf Right (b):} Electron energy spectra due to various DM signal points and principal beam-on backgrounds (unstacked histograms) assuming the on-axis \dae/LENA configuration.  The color shaded region under each signal curve represents the signal window that maximizes $S/\delta B$ for each parameter point. The $\nu_\mu$ CCQE distribution shows the residual background after 
a 70\% reduction from vetoing Michel electrons; the remaining muons are mis-identified as electrons in LENA, and their kinetic energy spectrum is shown. The $\nu_e$ CCQE distribution was only simulated above 100 MeV where it begins to dominate. The $\epsilon^2$ values for each signal point are chosen to match the minimum value for which the \dae/LENA setup has the $3\sigma$ sensitivity displayed in Fig. \ref{fig:money}.  
}
\label{fig:SigvsBGDistributions} 
\vspace{0cm}
\end{figure}

%----------------------------------------
%Beam-on table
%----------------------------------------

\begin{table*}[tdp]
\begin{center}\begin{tabular}{|c||c|c|c|c|c|c|}
\hline \hline
\textbf{Source} & \textbf{Neutrino} &  \textbf{Reaction Type}  & \textbf{106--147 MeV} & \textbf{147--250 MeV} & \textbf{250--400 MeV} & \textbf{Tag}  \\
\hline
\hline
\multirow{4}{*}{$\pi^+$ DIF}
& \multirow{2}{*}{$\nu_\mu$}
& elastic & \textbf{959} & \textbf{316} & $\hspace{-0.3cm}< 1$ & --\\
\cline{3-7}
& & CCQE & \textbf{1650} & 0 & 0 & Michel \\
\cline{2-7}
& \multirow{2}{*}{$\nu_e$}
& elastic & 4 & 5 & 2 & -- \\
\cline{3-7}
& & CCQE & 65 & \textbf{214} & \textbf{331} & --\\
\hline
\multirow{4}{*}{$\pi^-$ DIF}
& \multirow{2}{*}{$\overline{\nu}_\mu$}
&  elastic & 130 & 42 & \hspace{-0.3cm}$<1$ & -- \\
\cline{3-7}
& & CCQE & 382 & 0 & 0 & Michel \\
\cline{2-7}
& \multirow{2}{*}{$\overline{\nu}_e$}
& elastic &  \hspace{-0.3cm}$<1$ &  \hspace{-0.3cm}$<1$ &  \hspace{-0.3cm}$<1$ & -- \\
\cline{3-7}
& & CCQE & 7 &  23 &  36 & neutron \\
\hline
\hline
\end{tabular}
\caption{One-year rates for all beam-on backgrounds resulting in an outgoing lepton $\ell = e, \mu$ with kinetic energy $T_\ell > 106 \ \MeV$ in the final state. ``Elastic'' refers to elastic neutrino-electron scattering, and ``CCQE'' refers to charged-current quasi-elastic neutrino-nucleon scattering. A cut $\cos \theta_\ell > 0.9$ has been imposed on all outgoing charged leptons. Bolded entries are dominant backgrounds in their respective energy ranges. We expect backgrounds from $\mu^+$ decay-in-flight (DIF) to be subdominant; see text for details.}
\label{tab:AllBeamOnBackgrounds}
\end{center}
\end{table*}

We now consider the possible beam-on backgrounds. By imposing kinematic cuts which select for neutrino energies $E_\nu > 52.8 \ \MeV$, we eliminate the large decay-at-rest neutrino background from
\be
\pi^+ \to \mu^+ \nu_\mu, \qquad \mu^+ \to e^+ \nu_e \overline{\nu}_\mu.
\label{eq:PiDAR}
\ee
A further cut at $E_\nu > 70 \ \MeV$ eliminates the neutrino background from helicity-suppressed $\pi^+$ decays-at-rest,
\be
\pi^+ \to e^+ \nu_e,
\ee
which could pose a significant background because of the large number of stopped pions at \dae. Finally, a cut at $E_\nu > m_\mu \approx 106 \ \MeV$ mitigates the neutrino background from muon capture,
\be
\mu^- + {}^{A}_{Z}{N} \to \nu_\mu + {}^{A}_{Z-1}{N}',
\ee
where $ N$ is a nucleus in the \dae\ target, either carbon or copper. The rate of muon capture is not well-modeled by our GEANT simulation since the true \dae\ target contains copper, and the cross section for $\mu^-$ capture on copper is much higher than on graphite. However, the neutrinos produced from muon capture have a sharp kinematic endpoint at or below the muon mass, and suffer an acceptance penalty because they are produced isotropically, so we expect this background to be negligible above 106 MeV.

The remaining beam-on sources of neutrinos above 106 MeV are all decays-in-flight,
\begin{align}
\pi^+ & \to \mu^+ \nu_\mu, \\
\pi^+ & \to e^+ \nu_e, \\
\pi^- & \to \mu^- \overline{\nu}_\mu,\\
\pi^- & \to e^- \overline{\nu}_e, \\
\mu^+ & \to e^+ \, \overline{\nu}_\mu \, \nu_e.
\end{align}
Note the inclusion of the helicity-suppressed pion decay modes to electrons and positrons, which will in fact pose the main backgrounds above 250 MeV. To estimate the beam-on backgrounds, we used the same GEANT simulation which generated our signal events to generate the parent pions and muons, and GENIE \cite{Andreopoulos:2009rq} to simulate the CCQE processes; details are given in \App{app:NeutrinoBGBeamOn}.  The simplified \dae\ target geometry used in this simulation consisted of a single block of graphite with a flat face, whereas the full \dae\ design consists of a graphite and copper target with a re-entrant hole.  Since the stopping power for copper is greater than for graphite, we expect the decay-in-flight background from this simulation to be an upper limit on the true decay-in-flight background from the \dae\ neutrino source.  Furthermore, we expect our simulation to over-estimate the number of backscattered pions, since in the full \dae\ target design, some pions will stop in target material surrounding the re-entrant hole.  That said, since we focus on energies above the decay-at-rest neutrino spectrum, these backscattered pions do not pose a background in this analysis. 

A few words are in order regarding our treatment of the muon decay-in-flight backgrounds. For the LSND experiment, the $\nu_e$ background from $\mu^+$ decays was of the same order of magnitude as that from $\pi^+$ decays, in the electron recoil range 60--200 MeV \cite{Athanassopoulos:1997er}. However, at \dae, we expect the $\nu_e$ background from $\pi^+$ decay to be dominant for a number of reasons.  First, a significant number of the decay-in-flight $\mu^+$ at LSND were due to isotope stringers placed in the LAMPF beam upstream of the LSND target, whereas the \dae\ target will be optimized to suppress decay-in-flight backgrounds.  Second, the spectrum of decay-in-flight $\mu^+$ at \dae\ is much softer than the $\pi^+$ spectrum due to the longer muon lifetime and correspondingly larger energy loss in the \dae\ target. Third, the daughter neutrinos are less energetic: 52.4 MeV in the muon rest frame, as compared to 70 MeV in the pion rest frame. Therefore we expect this background to be subdominant to the $\pi^+$ decay-in-flight $\nu_e$ CCQE background for energies above 250 MeV, and subdominant to the $\pi^+$ decay-in-flight $\nu_\mu$-electron elastic scattering background between 106 and 250 MeV. We attempted to directly simulate this background with GEANT, but statistics proved prohibitive; we leave a full simulation of this background to more detailed studies.

Exactly as with beam-off backgrounds, beam-on backgrounds consist of both $\nu-e^-$ elastic scattering and CCQE events. Elastic events tend to have the outgoing electron scattered at small angles with respect to the initial neutrino direction when $T_e > 106 \ \MeV$, while CCQE events tend to have the lepton (electron or muon) produced more isotropically. As shown in \Fig{fig:SigvsBGDistributions:a}, the DM distribution is strongly peaked in the forward direction, such that much of the signal at large recoil energies will have electrons nearly parallel to the beamline.\footnote{The fact that the beam-on neutrino angular distribution appears to rise in the backwards direction is an artifact of our simplified GEANT simulation; without a re-entrant hole, we have a large number of backscattered pions.} Thus beam-on CCQE background events can be mitigated with the same cut on the outgoing charged lepton angle $\theta_\ell < 25^\circ$ as was used for beam-off events. The uncertainty for beam-on backgrounds is dominated by the systematic uncertainty in the neutrino flux. For each flavor of neutrino, a charged-current (CC) channel is available to measure the flux:
\begin{align}
\label{eq:CCQENuMu}
\nu_\mu \ {}^{12}{\rm C}  \to \mu^- \ X \qquad & \textrm{(tagged muon)},\\
\label{eq:CCQENuE}
\nu_e \ {}^{12}{\rm C}  \to e^-  \ {}^{12}N_{\rm gs} \qquad & \textrm{(tagged}\ ^{12}N_{\rm gs} \textrm{ beta decay)},\\
\label{eq:CCQEAntiNuMu}
\overline{\nu}_\mu \, p  \to \mu^+ n  \qquad & \textrm{(tagged muon and neutron)},\\
\label{eq:CCQEAntiNuE}
\overline{\nu}_e \, p  \to e^+ n  \qquad & \textrm{(tagged neutron)}.
\vspace{3cm}
\end{align}
There has been a considerable experimental effort to measure these CC cross sections \cite{Formaggio:2013kya}, and recently it was proposed to measure the inclusive CC reaction in \Eq{eq:CCQENuMu} with a mono-energetic 236 \MeV\ $\nu_\mu$ beam from kaon decays \cite{Spitz:2014hwa}.  When presenting the reach of \dae/LENA, we will assume a 20\% uncertainty in all of these cross sections, translating to an approximate 20\% uncertainty in all beam-on background rates, $\delta B = 0.2 B$.\footnote{The high statistics of the JPARC-MLF experiment \cite{Harada:2013yaa}, which should see nearly 200,000 CCQE events at 236 MeV, would give a much better than 20\% uncertainty on the differential energy spectrum. However, there would still be considerable uncertainty on the overall normalization, since theoretical predictions for the inclusive CC cross section can differ up to 25\% (see \Ref{Spitz:2014hwa} for a discussion). That said, the exclusive channel in \Eq{eq:CCQENuE}, which accounts for about 1\% of the $\nu_e$ CCQE cross section, has a smaller $\simeq 10\%$ uncertainty and may be useful for determination of the absolute flux to 10\%. We thank Joshua Spitz for bringing this point to our attention.}

The elastic and CCQE rates for all beam-on backgrounds above 106 MeV with the angular cut imposed are summarized in \Tab{tab:AllBeamOnBackgrounds} for the three benchmark energy ranges. The main irreducible background in the recoil energy range 106--147 MeV is $\nu_\mu-e^-$ elastic scattering. The main reducible background is $\nu_\mu$ CCQE, which produces an outgoing muon; 70\% of the time this muon can be identified through its Michel electron decay product~\cite{Wurm:2011zn}, which as described above also provides the channel with which to calibrate the $\nu_\mu$ flux. Above 147 MeV, muons can no longer be produced in CCQE events from beam-on neutrino sources, leaving the $\nu_\mu-e^-$ elastic background as the dominant irreducible background in the recoil energy range 147--250 MeV, with a significant contribution from $\nu_e$ CCQE. Above 250 MeV, the rate due to beam-on $\nu_\mu-e^-$ elastic scattering is less than 1 event per year. Here, the dominant background is $\nu_e$ CCQE. Amusingly, the source of these electron neutrinos is the helicity-suppressed decay $\pi^+ \to e^+ \nu_e$, which despite its branching ratio of $1.23 \times 10^{-4}$, has a very broad $\nu_e$ energy spectrum and a large CCQE cross section. The corresponding decay $\pi^- \to e^-  \overline{\nu}_e$ leads to a subdominant reducible background with a taggable neutron.

\begin{figure}[t!] 
 \includegraphics[width=9cm]{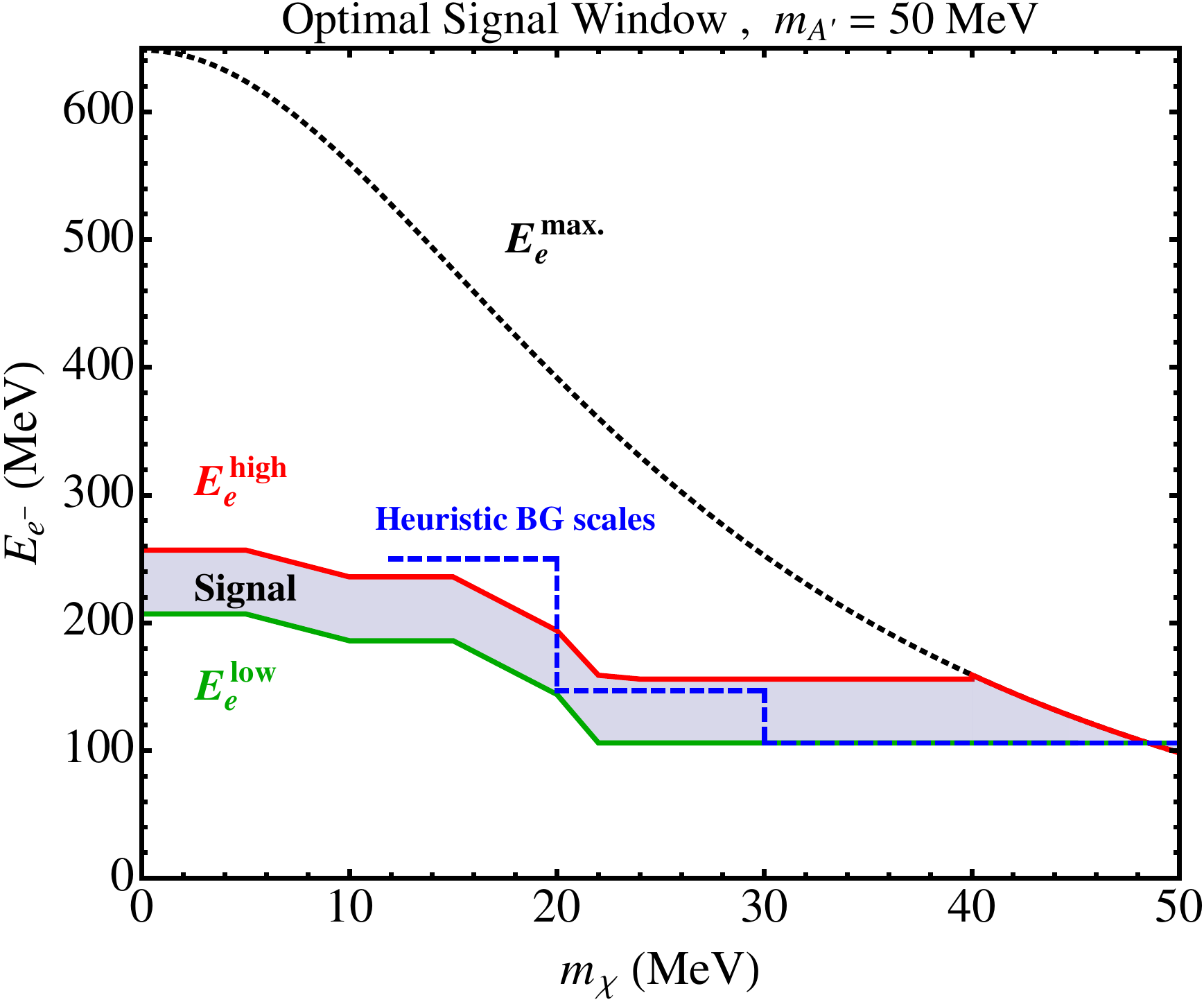} 
 \caption{Optimal electron recoil cuts $E^{\rm low}_e$ (green curve) and $E^{\rm high}_e$ (red curve), which optimize the signal-to-background sensitivity $S/\delta B$ as a function of $m_\chi$ for 
 fixed $m_{A'} = 50 $ MeV, assuming a minimum signal window width of 50 MeV. The shaded region between the red and green curves defines the optimal signal window for each mass point. Also shown is the 
 maximum electron recoil energy $E^{\rm max.}_e$ (black, dotted curve) for each $m_\chi$ assuming an initial proton energy 
 of 800 MeV (see Eq. \ref{eq:maxEEnergy}). The blue dashed lines at $E_e = 106, 147,$ and $ 250$ MeV respectively denote the electron energies beyond which 
beam-on backgrounds from $\mu^-$ capture, $\nu_\mu$  CCQE (from $\pi^+$ DIF),  and $\nu_\mu$ elastic scattering (from  $\pi^+$ DIF) become irrelevant; these lines can be regarded as a heuristic estimate of $E_e^{\rm low}(m_\chi)$. Above 250 MeV, the only significant beam-on background is from the $\nu_e$ CCQE process (see Table \ref{tab:AllBeamOnBackgrounds}). 
   }\label{fig:OptimalCuts}
\end{figure}

The optimal recoil cuts as a function of $m_\chi$ and $m_{A'}$ can now be determined based on the various background thresholds. For light $\chi$, \Fig{fig:SigvsBGDistributions:b} shows that the DM recoil spectrum is relatively flat and extends to high energies, so the optimal $E_e^{\rm low}$ is around 210 MeV where the only significant background is $\nu_e$ CCQE. As $m_\chi$ increases, the DM distribution begins to fall more steeply with energy, such that for $m_\chi \simeq 20 \ \MeV$ the signal and $\nu_\mu$ elastic background fall at approximately the same rate. Thus, one needs to apply a lower energy cut to retain a sufficient yield of signal events; this is true for both on- and off-shell DM production. Below 250 MeV the only new background is $\nu_\mu$ elastic scattering, so to keep the maximum number of signal events, the optimal $E_e^{\rm low}$ should be close to 147 MeV. For heavier DM, $m_\chi \gtrsim 30 \ \MeV$, the 147 MeV cut is too severe because the DM is not produced with enough kinetic energy to provoke recoils above 147 MeV at an appreciable rate. As described above, to avoid the numerous low-energy backgrounds, the lowest realistic energy cut is $E_e^{\rm low} = 106 \ \MeV$. We determined $E_e^{\rm high}$ as a function of $m_\chi$ and $m_{A'}$ by optimizing signal-to-background sensitivity $S/\delta B$ using $\delta B = \sqrt{4B/3}$ (systematic and statistical errors combined) for beam-off and $\delta B = 0.2 B$ (systematic only) for beam-on; the result for $m_{A'} = 50 \ \MeV$ is shown in \Fig{fig:OptimalCuts}. Due to the broad neutrino background spectra, the optimal signal window is as narrow as possible for all DM masses. However, the energy resolution at LENA is on the order of a few percent in the energy range we consider \cite{Wurm:2011zn}. To be conservative, we use signal windows of 50 MeV or greater in electron recoil energy.

% -----------------------------------------------------------------------------------------------------------------------------------------------------------------
%						Optimal recoil cuts 
% -----------------------------------------------------------------------------------------------------------------------------------------------------------------

% -----------------------------------------------------------------------------------------------------------------------------------------------------------------

\section{Sensitivity}
\label{sec:Sensitivity}

\begin{figure}[t!] 
 \includegraphics[width=9cm]{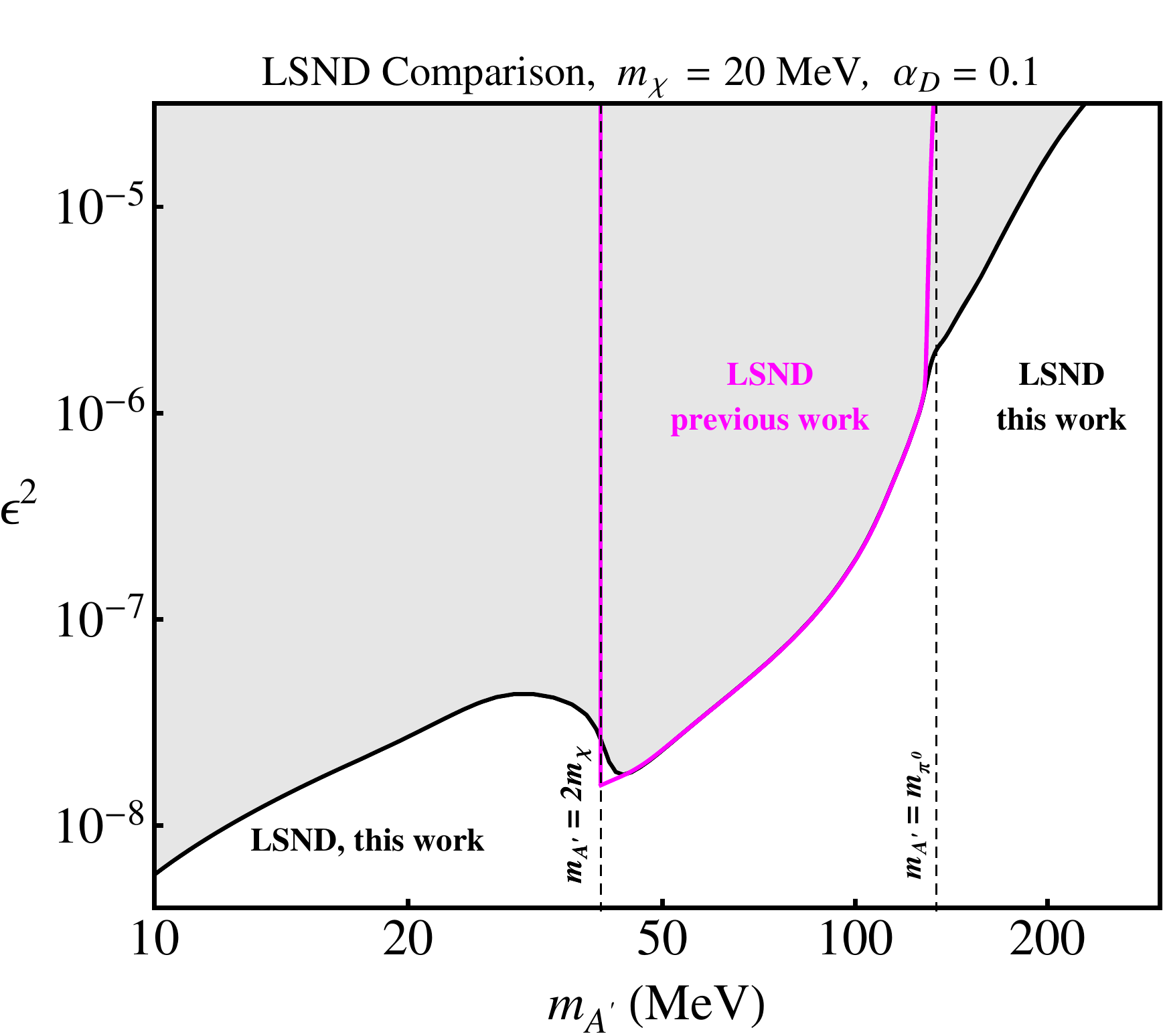} 
 \caption{Comparison of LSND sensitivities as computed using methods in the existing literature \cite{Batell:2009di, deNiverville:2011it} (magenta curve) and those obtained using the  full three-body matrix element that includes DM production via an off-shell $A'$. \label{fig:LSNDcomparison} }
\end{figure}

The main results of this paper are shown in \Fig{fig:money}, which give the $3\sigma$ sensitivity of the \dae/LENA setup to the dark photon/DM parameter space.  We also show updated results for the LSND exclusions, which extend the analysis of \Ref{deNiverville:2011it} into both off-shell $A'$ regimes.  Our LSND exclusions are based on rescaling our GEANT simulation for the DM signal rates in \dae/LENA to match the collision rate and target geometry of LSND. We make no attempt to simulate the backgrounds at LSND, but instead assume that the 55-event upper limit quoted in \Ref{Auerbach:2001wg} accounts for background subtraction. Our signal yields are expected to be very similar to the analysis in \Ref{deNiverville:2011it}, because the $\pi^0$ spectrum depends very little on the target geometry; we verified that in the on-shell $A'$ regime, we obtain nearly identical results to \Ref{deNiverville:2011it}.  A key feature to note is the dark gray bands in \Figs{fig:money:c}{fig:money:e}, which indicate the region of parameter space where LSND can place bounds on \emph{visible} $A' \to e^+ e^-$ decays by searching for DM produced in $\pi^0 \to \gamma A^{\prime*} \to \gamma \chi \overline{\chi}$ via an off-shell $A^{\prime}$.  The extended exclusion limits from LSND compared to the previously-reported limits are demonstrated in \Fig{fig:LSNDcomparison} for $m_\chi = 20 \ \MeV$; we discuss the reason for this extended coverage in more detail below.

Also plotted in \Fig{fig:money} are constraints and projected sensitivities for a variety of dark photon searches; for a comprehensive review of this parameter space see \Ref{Essig:2013lka} and citations therein.  The constraints are from E137 \cite{Bjorken:1988as, Batell:2014mga}, Orsay \cite{Davier:1989wz}, muon $g-2$ \cite{Pospelov:2008zw,Endo:2012hp}, electron $g-2$ \cite{Giudice:2012ms, Hanneke:2008tm}, E141 \cite{Bjorken:2009mm},  E787 \cite{Adler:2004hp}, E949 \cite{Artamonov:2009sz}, the BaBar visible search for $A' \to e^+e^-$ \cite{Lees:2014xha} denoted ``BaBar V" in \Fig{fig:money}, the BaBar invisible search for monophoton and missing energy \cite{Aubert:2008as} denoted ``BaBar I" in \Fig{fig:money},  and preliminary results from NA48/2 \cite{NA48}.  
Other visible constraints from A1 \cite{Merkel:2011ze}, and the  APEX test run \cite{Abrahamyan:2011gv}  are shown in Fig. \ref{fig:g2favored}; recent constraints from PHENIX \cite{Adare:2014mgk} are subdominant to NA48/2 in this region of parameter space. The projected sensitivities involve a combination of visible $A' \to e^+ e^-$ and invisible $A' \to \chi \overline{\chi}$ searches:  BDX \cite{Battaglieri:2014qoa},  APEX \cite{Essig:2010xa, Abrahamyan:2011gv},  HPS \cite{Moreno:2013mja}, MESA and MAMI \cite{Beranek:2013yqa}, VEPP-3 \cite{Wojtsekhowski:2012zq}, and DarkLight \cite{Freytsis:2009bh, Balewski:2013oza,Kahn:2012br}.  The thick green band is the parameter space for which $\apr$ resolves the  long-standing $(g-2)_\mu$ anomaly \cite{Pospelov:2008zw}.          
% removed reference to,  

% -----------------------------------------------------------------------------------------------------------------------------------------------------------------
%						LSND Comparison 
% -----------------------------------------------------------------------------------------------------------------------------------------------------------------

% -----------------------------------------------------------------------------------------------------------------------------------------------------------------

The plots in the left column of \Fig{fig:money} show the \dae/LENA sensitivity in $\epsilon^2$ for fixed $(\alpha_D, m_\chi)$ as a function of $m_{A'}$, where for each point $(m_{A'}, m_\chi)$ the signal window is chosen to optimize the sensitivity, as in \Fig{fig:OptimalCuts}. For light $\chi$ ($m_{\chi} = 1 \ \MeV$ in \Fig{fig:money:a}), the sensitivity curve is essentially parallel to that of LSND, but better by an order of magnitude due to the optimized cuts. The projected sensitivity of the BDX experiment is shown in dashed green for comparison. For this DM mass, the $A'$ is produced on-shell for $m_{A'} < m_{\pizero}$, and off-shell when $m_{A'} > m_{\pizero}$. However, there is no sharp kinematic threshold at $m_{A'} = m_{\pizero}$, and both \dae/LENA and LSND still have sensitivity in the upper off-shell regime; this observation was neglected in previous studies, due to an improper application of the narrow-width approximation.

Going to heavier DM, $m_\chi = 20 \ \MeV$ in \Fig{fig:money:c}, we can probe the on-shell region $2m_\chi < m_{A'} < m_{\pizero}$, as well as the two off-shell regions $m_{A'} < 2m_\chi$ and $m_{A'} > m_{\pizero}$.   The large mass of DM compared to the $A'$ results in two key differences compared to the light DM case.  First, there is a true kinematic threshold for on-shell production of the $A'$ at $m_{\apr} = 40 \ \MeV$. Just above threshold, the phase space suppression of DM particles produced nearly at rest in the on-shell $A'$ rest frame competes with the matrix element suppression of DM produced through an off-shell $A'$, and so the cut on electron recoil energy tends to shift the point of maximum sensitivity in $\epsilon^2$ to larger $A'$ masses. This results in a dip at $m_{\apr} \gtrsim 40 \ \MeV$ rather than a sharp drop exactly at threshold. Second, in the lower off-shell regime $m_{A'} < 40 \ \MeV$, both \dae\ and LSND are still sensitive to DM production and scattering, and in fact the sensitivity to very light off-shell $A'$s is superior to the on-shell sensitivity.  This surprising observation has also been neglected in previous studies, and is possible because the virtuality of the $A'$ does not generate all that much of a suppression in the decay $\pi^0 \to \gamma A^{\prime*} \to \gamma \chi \overline{\chi}$.  Indeed, phase-space constraints at high $m_A'$ can be more restrictive than matrix element suppression at low $m_A'$, such that there is a region of parameter space at very low $m_A'$ where the off-shell reach of both experiments in $epsilon^2$ is stronger than the on-shell reach.

Furthermore, because the $A'$ couples to electrons by assumption, if $A'$ decays to DM are kinematically forbidden, then the decay channel $A' \to e^+ e^-$ must be open.  This leads to the key feature mentioned above that the sensitivity of \dae/LENA and LSND in the lower off-shell $A'$ regime can overlap with visible $A' \to e^+ e^-$ searches.  Indeed, for $m_\chi = 20 \ \MeV$, the reach of LSND and \dae/LENA is comparable to experiments like E141 \cite{Bjorken:2009mm} and HPS \cite{Moreno:2013mja}.  Of course, the visible limits are independent of $m_\chi$ whereas the LSND and \dae/LENA limits require a dark sector state of the appropriate mass.  Still, this emphasizes the importance of studying the full $\{ m_{A'}, \epsilon, m_\chi, \alpha_D \}$ parameter space.  Note that as $\alpha_D$ increases, the LSND and \dae\  curves on these plots shift downward. DM production is independent of $\alpha_D$ in the on-shell regime but proportional to $\alpha_D$ in the off-shell regime, while DM scattering is proportional to $\alpha_D$ for any $m_\chi$ and $m_{A'}$ (see \App{app:Production} and \App{app:Scattering}). Thus, the scaling of the sensitivity with $\alpha_D$ is quadratic in the off-shell regime and linear in the on-shell regime. In contrast, the visible searches remain unaffected as $\alpha_D$ is changed since the on-shell $\apr \to e^+e^-$ process is independent of the dark coupling $\alpha_D$.

Going to even heavier DM, $m_\chi = 40 \ \MeV$ in \Fig{fig:money:e}, we see that constraints from LSND data already cover the entire region which would be probed by \dae\ in one year of running.  This is due to the fact that LSND is a \u{C}erenkov detector and can use directionality to discriminate against neutrino backgrounds at lower energies than LENA. 
For the \dae/LENA setup, the minimum recoil cut of 106 MeV which is necessary to mitigate the backgrounds also cuts out the majority of the signal, since the heavy DM is produced with relatively low kinetic energy. This also results in an even greater degradation of sensitivity near the on-shell threshold at $m_{A'} = 2m_\chi$ compared to LSND. Thus we see that experiments like LSND, which have sensitivity to low electron recoil energies, are optimal for larger $m_\chi$. 

The plots in the right column of \Fig{fig:money} show the sensitivity in $\epsilon^2$ for fixed $(m_{A'}, \alpha_D)$ as a function of $m_\chi$, where again the electron recoil cuts are chosen for each $m_\chi$ to optimize the sensitivity as in \Fig{fig:OptimalCuts}. The \dae/LENA reach improves on LSND by an order of magnitude for light $\chi$, but the improvement weakens for heavier $\chi$ for the same reasons discussed above: the LSND recoil cuts favor heavier DM because it is produced with less kinetic energy.  The constraints from visible searches now appear as horizontal lines in the off-shell regime because they depend only on $m_{A'}$ and not on $m_\chi$. 

Finally, \Fig{fig:g2favored} shows a different slice through parameter space.  Here, we fix $m_\chi$ and show the sensitivity to $\alpha_D$ as a function of $m_{A'}$, where for each $A'$ mass, $\epsilon$ assumes the lowest value consistent with the $(g-2)_\mu$ preferred band (as shown in green in \Fig{fig:money:a}).  We see that \dae/LENA can improve considerably on LSND bounds over the entire kinematically-accessible parameter space of the dark photon model, and nearly covers all of the remaining parameter space that resolves the $(g-2)_\mu$ anomaly. The prospect of reconciling this anomaly with a dark photon is usually discussed for an $A'$ which decays purely to $e^+e^-$ or purely to dark-sector states (see \Ref{Lee:2014tba} for a discussion of current constraints), but presenting the parameter space in this fashion shows that \dae/LENA is sensitive to dark photons that decay predominantly to visible states, and that visible decay experiments already cover some regions in which the $A'$ decays invisibly.\footnote{We thank Natalia Toro for pointing out the sensitivity of visible searches in this region of parameter space.} Note that after including preliminary results from NA48/2 \cite{NA48}, the $(g-2)_\mu$ window for a visibly decaying
$A'$ is now fully closed (see also Fig. \ref{fig:money:e}). 
\section{Conclusion}
\label{sec:Conclusion}

A rich dark sector remains a well-motivated possibility, and light DM coupled to a kinetically-mixed dark photon provides excellent opportunities for discovery. In this paper we have shown that intensity frontier experiments like \dae, in conjunction with a large underground neutrino detector such as LENA, will have unprecedented sensitivity to light (sub-50 MeV) DM, light (sub-400 MeV) dark photons, and other light, weakly coupled particles. Previous analyses have emphasized the $m_{\apr} > 2m_\chi$ region of parameter space where the $A^\prime$ decays almost exclusively to the dark sector via $\apr \to \chi \overline{\chi}$.  This focus was motivated by the typical size of $\epsilon$, which ensures that if light dark-sector states exist, then $\Br(A^\prime \to \chi \overline{\chi}) \approx 1$.  Here, we have shown that existing LSND data places strong constraints on two additional regions: the $m_{\apr} < 2m_\chi$ regime, where on-shell $\apr$s decay via the visible channel $A' \to e^+ e^-$ but DM can be produced via an off-shell $\apr$, and the $m_{\apr} > m_{\pizero} > 2m_\chi$ regime, which does not actually contain a kinematic threshold forbidding DM production. Because DM can be produced through both on- and off-shell dark photons, the full four-dimensional parameter space $\{ m_{A'}, \epsilon, m_\chi, \alpha_D \}$ contains interesting regimes which are not captured in the usual $\{ m_{A'}, \epsilon \}$ plots.  \dae\ is uniquely sensitive to this larger parameter space, even up to $A'$ masses of 500 MeV. We also encourage the current search at MiniBooNE to explore this expanded parameter space.

In addition to the potential advantages of higher luminosity and larger acceptance compared to previous experiments, a light DM search at \dae/LENA would not require a separate running mode, such as the off-target mode used for MiniBooNE. While the sensitivity is best in the on-axis configuration, the reach is relatively insensitive to the detector geometry, and so a DM search could run simultaneously with a decay-at-rest neutrino experiment, provided analysis cuts are performed offline after data-taking. In fact, pairing \dae\ with a large-volume underground \u{C}erenkov detector like the proposed Hyper-K, with sensitivity to both low and high electron recoil energies and good electron-muon separation to reduce CCQE backgrounds, could cover a broad region of the full four-dimensional parameter space of the dark photon model. The fact that both neutrino and DM experiments share essentially the same signals and backgrounds, though often well-separated kinematically, is an advantageous feature of such a setup, and suggests exciting opportunities for symbiosis between beyond-the-standard-model and neutrino physics in the coming years.

\acknowledgments{
We thank Adam Anderson,  Brian Batell, Janet Conrad, Rouven Essig, Joseph Formaggio, Eder Izaguirre, Patrick de Nieverville,  Maxim Pospelov, Philip Schuster, Joshua Spitz, and Natalia Toro for many helpful conversations. 
GK thanks the Aspen Center for Physics, supported by NSF Grant No. PHY-1066293, and the CERN theory group for hospitality while portions of this work were completed. 
The Perimeter Institute for Theoretical Physics is supported by the Government of Canada through Industry Canada and by the Province of Ontario. 
Y.K. and J.T. are supported by the U.S. Department of Energy (DOE) under grant Contract Number DE-SC00012567.  Y.K. is also supported by an NSF Graduate Research Fellowship.  J.T. is also supported by the DOE Early Career research program DE-FG02-11ER-41741 and by a Sloan Research Fellowship from the Alfred P. Sloan Foundation. M.T. is supported by the National Science Foundation under Grant Number PHY-1205175.}

\appendix

\section{Dark Matter Production Rates}
\label{app:Production}

For calculating the DM production rates and kinematics at \dae\ in \Sec{sec:Production}, we need the three-body matrix element for $\pi^0 \to \gamma A^{\prime(*)} \to \gamma \chi \overline{\chi}$, summed over photon polarizations and DM spins if necessary.  The calculations below are sufficiently general to be used for either an on-shell or off-shell $A'$, so we will keep the width $\Gamma_{A'}$ in the $A'$ propagator.  We will give expressions both for complex scalar DM and Dirac fermion DM, though we only show plots for fermionic DM in the text.

\subsection{Dark Photon Width}

For the parameter space $m_{A'} > 2m_e$ and assuming that $\chi$ is the only dark-sector particle coupled to U(1)$_D$, the $A'$ width is 
\begin{equation}
\Gamma_{A', \text{tot}} =
	\begin{cases}
	\Gamma_{A' \to \chi \overline{\chi}} + \Gamma_{A' \to e^+ e^-} &  (m_{A'} > 2m_\chi), \\
	\Gamma_{A' \to e^+ e^-} & (m_{A'} < 2m_\chi).
	\end{cases}
\end{equation} 
The two-body widths are given by
\begin{equation}
\Gamma_{A' \to X \overline{X}} = \frac{| \mathbf{p} |}{8 \pi m_{A'}^2} \langle |\mathcal{A}|^2 \rangle,
\label{eq:2bodytotalwidth}
\end{equation}
with $|\mathbf{p}| = \sqrt{m_{A'}^2/4-m_X^2}$, and $m_X = m_\chi$ or $m_e$ as appropriate. The spin-averaged squared amplitudes for $A'$ decay to DM and leptons are
\begin{align}
\langle |\mathcal{A}_{A' \to \chi \overline{\chi}}|^2 \rangle & = \frac{g_D^2}{3} \times
\begin{cases}
m_{A'}^2  - 4m_\chi^2 & \textrm{(scalar)}, \\
4 m_{A'}^2 + 8m_\chi^2 &  \textrm{(fermion)}, \\
\end{cases} \\
\langle |\mathcal{A}_{A' \to e^+ e^-}|^2\rangle & = \frac{4}{3}\epsilon^2 e^2(2m_e^2 + m_{A'}^2),
\end{align}
where $g_D$ is the U(1)$_D$ gauge coupling. The total $A'$ width is therefore
\begin{equation}
\label{eq:AprTotWidth}
\Gamma_{A', \text{tot}} = \frac{1}{6 m_{A'}^2} \times 
\begin{cases}
\alpha_D( m_{A'}^2  - 4m_\chi^2 ) \sqrt{m_{A'}^2/4-m_\chi^2} + 4\epsilon^2 \alphaEM (2m_e^2 + m_{A'}^2) \sqrt{m_{A'}^2/4-m_e^2} &\textrm{ (scalar)}, \\
4\alpha_D(m_{A'}^2 + 2m_\chi^2) \sqrt{m_{A'}^2/4-m_\chi^2} + 4\epsilon^2 \alphaEM(2m_e^2 + m_{A'}^2) \sqrt{m_{A'}^2/4-m_e^2} & \textrm{ (fermion)}, \\
4\epsilon^2 \alphaEM(2m_e^2 + m_{A'}^2) \sqrt{m_{A'}^2/4-m_e^2} & \textrm{ (off-shell)},
\end{cases}
\end{equation}
where $\alpha_D \equiv g_D^2/4\pi$ and $\alphaEM \equiv e^2/4\pi$ are the U(1)$_D$ and electromagnetic fine structure constants, respectively. The last expression is valid when $m_{A'} < 2m_\chi$ such that on-shell decays $A' \to \chi \overline{\chi}$ are kinematically forbidden.

\subsection{Scalar DM Production}
The matrix element for DM production can be obtained by replacing a photon leg with an $A'$ leg in the $\pizero \to \gamma \gamma$ effective vertex mediated by the chiral anomaly, with the $A' \to \chi \overline{\chi}$ part of the diagram determined by the U(1)$_D$ coupling to $\chi$. For the case of scalar DM, the matrix element is
\begin{equation}
\mathcal{A}_{\pi^0 \to \gamma \chi \overline{\chi}}  =\epsilon g_D \frac{e^2}{4\pi^2} \frac{1}{f_\pi} \epsilon_\lambda^{(\gamma)} \epsilon^{\lambda \mu \alpha \beta} p_\alpha q_\beta \frac{-i(g_{\mu \nu} - q_\mu q_\nu /m_{A'}^2)}{s - m_{A'}^2 + i m_{A'} \Gamma_{A'}}(k_2^\nu - k_1^\nu )  \hspace{3mm} \textrm{(scalar)},
\end{equation}
where $p$ is the photon momentum, $k_1$ and $k_2$ are the DM momenta, $q = k_1 + k_2$ is the virtual $A'$ momentum, $s = q^2$, $\epsilon_\lambda^{(\gamma)}$ is the polarization vector of the outgoing photon, and $f_\pi$ is the pion decay constant. Squaring and summing over the two photon polarizations gives
\begin{equation}
\langle |\mathcal{A}_{\pizero \to \gamma  \chi \overline{\chi}} |^2 \rangle = -\frac{\epsilon^2 g_D^2 \alphaEM^2}{\pi^2 f_\pi^2} g_{\lambda \rho} \epsilon^{\lambda \mu \alpha \beta} \epsilon^{\rho \sigma \gamma \delta} p_\alpha q_\beta p_\gamma q_\delta \frac{(g_{\mu \nu} - q_\mu q_\nu /m_{A'}^2)(g_{\sigma \kappa} - q_\sigma q_\kappa /m_{A'}^2)}{(s - m_{A'}^2)^2+m_{A'}^2 \Gamma_{A'}^2}(k_2^\nu - k_1^\nu )(k_2^\kappa - k_1^\kappa )  \hspace{3mm} \textrm{(scalar)}.
\end{equation}
There are six contractions of the $\epsilon$ tensors; two of them vanish identically because they result in a prefactor of $p^2 = 0$, and the remaining four can be simplified using $q\cdot(k_2 - k_1) = (k_2 + k_1)\cdot(k_2 - k_1) = k_2^2 - k_1^2 = 0$. This last identity ensures that all terms resulting from the $q_\mu q_\nu /m_{A'}^2$ part of the $A'$ propagator vanish, which must happen because the $A'$ couples to the conserved electromagnetic current.  We can also simplify some of the dot products using
\be
p \cdot q = \frac{m_\pi^2 - s}{2}, \qquad
k_1 \cdot k_2  = \frac{s}{2} - m_\chi^2,
\ee
which leads to the final result
\be
\label{DMProdScalar}
\langle |\mathcal{A}_{\pizero \to \gamma \chi \overline{\chi}}|^2 \rangle   = \frac{
\epsilon^2 \alphaEM^2 \alpha_D }{\pi f_\pi^2[(s - m_{A'}^2)^2+m_{A'}^2 \Gamma_{A'}^2]}   \left [         (s - 4m_{\chi}^2)\left (m_{\pizero}^2 -s\right)^2 -4 s( p \cdot k_1 - p \cdot k_2)^2 \right]  \hspace{3mm} \textrm{(scalar)}.
\ee
If $m_{A'} < 2m_\chi$, the $A'$ is off-shell, and the $A'$ width (which is proportional to $\epsilon^2$) can be neglected in the denominator; see \Eq{eq:AprTotWidth}.

\subsection{Fermionic DM Production}

The matrix element for fermionic DM is identical to the scalar case apart from the external spinors which replace the momentum factor $k_2^\nu - k_1^\nu$.  The matrix element is
\begin{equation}
\mathcal{A}_{\pi^0 \to \gamma \chi \overline{\chi}}  = \epsilon g_D \frac{e^2}{4\pi^2} \frac{1}{f_\pi} \epsilon_\lambda^{(\gamma)} \epsilon^{\lambda \mu \alpha \beta} p_\alpha q_\beta \frac{-i(g_{\mu \nu} - q_\mu q_\nu /m_{A'}^2)}{s - m_{A'}^2 + i m_{A'} \Gamma_{A'}} (\bar v(k_2) \gamma^\nu u(k_1))  \hspace{3mm} \textrm{(fermion)}.
\end{equation}
The additional spin sum is straightforward:
\begin{align}
\langle |\mathcal{A}_{\pizero \to \gamma \chi \chi}|^2 \rangle &= -\frac{4\epsilon^2 g_D^2 \alpha_{\rm EM}^2}{\pi^2 f_\pi^2} g_{\lambda \rho} \epsilon^{\lambda \mu \alpha \beta} \epsilon^{\rho \sigma \gamma \delta} p_\alpha q_\beta p_\gamma q_\delta \frac{(g_{\mu \nu} - q_\mu q_\nu /m_{A'}^2)(g_{\sigma \kappa} - q_\sigma q_\kappa /m_{A'}^2)}{(s - m_{A'}^2)^2+m_{A'}^2 \Gamma_{A'}^2} \nonumber \\
& \qquad \times [k_1^\nu k_2^\kappa + k_2^\nu k_1^\kappa -g^{\nu \kappa}(k_1 \cdot k_2 + m_\chi^2)]   \hspace{3mm} \textrm{(fermion)}.
\end{align}
The same two contractions as in the scalar case vanish from $p^2 = 0$, and indeed, the longitudinal part of the propagator still vanishes when contracted into the last term above. Simplifying this expression using the dot product identities above gives
\be
\label{DMProdFermion}
\langle |\mathcal{A}_{\pizero \to \gamma \chi \overline{\chi}}|^2 \rangle  = \frac{4\epsilon^2 \alphaEM^2 \alpha_D }{ \pi f_\pi^2[(s - m_{A'}^2)^2+m_{A'}^2 \Gamma_{A'}^2]}
 \left [(s + 2m_{\chi}^2)\left (m_{\pizero}^2 -s\right)^2 - 8s(p \cdot k_1)(p \cdot k_2) \right]   \hspace{3mm} \textrm{(fermion)}.
\ee
Again, if $m_{A'} < 2m_\chi$, the $A'$ width can be neglected.

\subsection{On-shell Regime}
\label{app:OnShell}

If the pole of the $\apr$ propagator is well within the physical kinematical region $s \in [4m_\chi^2, m_{\pizero}^2]$, we can use the narrow-width approximation \cite{SchwartzQFT},
\begin{equation}
\label{eq:NarrowWidthDelta}
\frac{1}{(s - m_{A'}^2)^2+m_{A'}^2 \Gamma_{A'}^2} \to \frac{\pi}{m_{A'} \Gamma_{A'}} \delta(s - m_{A'}^2).
\end{equation}
Making this substitution in the appropriate matrix elements and integrating over the phase space gives \Eq{BRDM} in the text. In particular, when $\alpha_D \gg \epsilon^2 \alphaEM$, $\Gamma_{A'} \propto \alpha_D$  (see \Eq{eq:AprTotWidth}), so the factors of $\alpha_D$ cancel and $\Gamma_{\pizero \to \gamma \chi \overline{\chi}}$ is independent of $\alpha_D$. However, if $\alpha_D \ll \epsilon^2 \alphaEM$ (as in a portion of parameter space that we consider in \Fig{fig:g2favored}), then $\Gamma_{A'} \propto \epsilon^2$ since the visible width dominates; in that case the factors of $\epsilon^2$ cancel and $\Gamma_{\pizero \to \gamma \chi \overline{\chi}}$ is proportional to $\alpha_D$ but independent of $\epsilon$.

As a check of the narrow-width approximation, we find the expected result 
\begin{equation}
\Gamma_{\pizero \to \gamma \chi \overline{\chi}} = \Gamma_{\pizero \to \gamma A'} \times \Br(A' \to \chi \overline{\chi}) \hspace{3mm} \textrm{(on-shell)},
\end{equation}
valid for both fermionic and scalar DM.

\subsection{Pion Threshold Regime}
\label{app:Threshold}

If the pole of the $\apr$ propagator is sufficiently close to $m_{\pizero}^2$, the narrow width approximation breaks down because the Breit-Wigner is no longer completely contained in the physical kinematical region $s \in [4m_\chi^2, m_{\pizero}^2]$. In that case, we must integrate the appropriate full three-body matrix element over phase space as in \Eq{BRDMOffshell} to obtain the branching ratio $\Br(\pizero \to \gamma \chi \overline{\chi})$. Now, however, the width must be included in the denominator because it is not parametrically small with our choice of parameters; it is proportional to $\alpha_D$ rather than $\epsilon^2$. In practice, the three-body matrix element must be used for $|m_{A'}^2 - m_{\pizero}^2| \lesssim 10 \Gamma_{A'} m_{A'}$; for $\alpha_D = 0.1$ and $m_\chi = 1 \ \MeV$, this translates to $120 \ \MeV \lesssim m_{A'} \lesssim 140 \ \MeV$.

 In the limit of large $m_{A'}$ and small $m_\chi$, the decay width for $\pizero \to \gamma \chi \overline{\chi}$ can be written as
\begin{equation}
\Gamma_{\pizero \to \gamma \chi \overline{\chi}} = \frac{m_{\pizero}^4}{120} \left(\frac{\epsilon g_D}{m_{A'}^2}\right)^2 \Gamma_{\pizero \to \gamma \gamma}.
\end{equation}
Thus we can view $\epsilon g_D/m_{A'}^2$ as a ``Fermi constant'' for the dark sector arising from integrating out the $A'$, analogous to integrating out the $W$ boson in the weak sector. This gives the scaling of the limits in the curves in \Figs{fig:money}{fig:LSNDcomparison} for $m_{A'} \gg m_{\pizero}$.
\section{Dark Matter Scattering Rates}
\label{app:Scattering}

For calculating the scattering of DM at LENA in \Sec{sec:Scattering}, we need to calculate the $\chi e^- \to \chi e^-$ differential cross section $d \sigma/ d E_e$.  While we have in mind elastic scattering off electrons, we will present formulas that are sufficiently general to apply to any point-like (fermionic) target $T$, and any inelastic splittings between DM masses which could lead to alternative signals.  We let the incoming (outgoing) DM have four-momentum $p_1$ ($k_1$) and mass $m_1$ ($m_2$). We assume the target $T$ is initially at rest in the lab frame, with mass $m_T$ and initial (final) four-momentum $p_2$ ($k_2$).  The case of $\chi e^- \to \chi e^-$ in the text is obtained with $m_1 = m_2 \equiv m_\chi$ and $T = e^-$.

\subsection{DM Scattering Amplitudes}

For scalar DM and a fermionic target $T$ (i.e.\ electron or nucleon), the amplitude for scattering via a $t$-channel kinetically mixed photon is
\be
{\cal A} = \frac{\epsilon e g_D }{(t - m^2_{A^\prime})} \bar u(k_2) (\displaystyle{\not}{p_1} + \displaystyle{\not}{k_1} ) u(p_2)  \hspace{3mm} \textrm{(scalar)}.
\ee
Unlike in the production case, here we can always ignore the $\apr$ width.  Squaring and averaging (summing) over the initial (final) state target spins gives
\begin{align}
\langle |{\cal A}|^2 \rangle
&= \frac{32\pi^2 \epsilon^2 \alphaEM \alpha_D }{(t - m^2_{A^\prime})^2} \biggl[
         (k_2\cdot p_1)(p_2\cdot p_1) + (k_2\cdot p_1)(p_2\cdot p_1) - (k_2\cdot p_2)(p_1\cdot p_1)    + 
        	(k_2\cdot p_1)(p_2\cdot k_1) + (k_2\cdot k_1)(p_2\cdot p_1) \nonumber \\ & \qquad\qquad \qquad ~ - (k_2\cdot p_2)(p_1\cdot k_1) +
         	(k_2\cdot k_1)(p_2\cdot k_1) + (k_2\cdot k_1)(p_2\cdot k_1) - (k_2\cdot p_2)(k_1\cdot k_1)   +         
         	(k_2\cdot k_1)(p_2\cdot p_1)\nonumber \\ & \qquad\qquad \qquad ~ + (k_2\cdot p_1)(p_2\cdot k_1)  - (k_2\cdot p_2)(k_1\cdot p_1) +   m_T^2 \left[     m_1^2  + m_2^2 +  2 (p_1\cdot k_1)  \right]  \biggr]   \hspace{3mm} \textrm{(scalar)},
\end{align}
where $t \equiv (k_1 - p_1)^2 = (k_2 - p_2)^2 = 2m_T^2 - 2 m_T E_{k_2}$ and $E_{k_2} = k^0_2$ in the lab frame. 
All quantities can now be written in terms of the incoming $\chi_1$ energy $E_{p_1}$ and the target 
recoil energy $E_{k_2}$ in the lab frame.

For fermionic DM, the analogous matrix element is 
\be
{\cal A} = \frac{\epsilon e g_D }{(t - m^2_{A^\prime})} [\bar u(k_2)\gamma_\mu u(p_2) ] [ \bar u(k_1)\gamma^\mu  u(p_1)]   \hspace{3mm} \textrm{(fermion)}.
\ee
Squaring and averaging/summing over the spin states gives
\begin{align}
\langle |{\cal A}|^2 \rangle
 &= \frac{128\pi^2 \epsilon^2 \alphaEM \alpha_D }{(t - m^2_{A^\prime})^2}  \left[    
(k_1\cdot k_2)(p_1\cdot p_2) + (k_2\cdot p_1)(p_2\cdot k_1) -  m_1 m_2 (k_2\cdot p_2) -  m_T^2 (p_1\cdot k_1) + 2 m_1 m_2 m_T^2  
\right]    \hspace{3mm} \textrm{(fermion)}.
\end{align}

\subsection{Differential Distributions}

From the amplitudes above, we can obtain the differential cross section.  Letting $^{*}$ denote quantities in the center-of-mass (CM) frame, the angular distribution is 
\be \label{eq:xsecCM}
\frac{d\sigma}{d\Omega^*} &=&
\frac{1}{2\pi}\frac{d\sigma}{d\cos\theta^*} = \frac{  \langle |{\cal A}|^2 \rangle}{  64 \pi^2 s  }    \frac{  |\vec k^*|  }{   \left|  \vec p^*  \right| },
\ee
where the initial and final state three-momenta in the CM frame are
 \be
  |\vec p^{\,*}|^2 =  \frac{ (s-m_T^2-m_1^2)^2 - 4 m_T^2 m_1^2}{4s}, \qquad  |\vec k^{\,*}|^2=  \frac{ (s-m_T^2-m_2^2)^2 - 4 m_T^2 m_2^2}{4s}.
 \ee
To go to the lab frame (without $^{*}$s), we can use the relations
\begin{align}
s &= (p_1 + p_2)^2 = m_1^2 + m_T^2 + 2m_T E_{p_1}, \\
k_1\cdot p_1 & = -\frac{1}{2} ( 2m_T^2 - m_1^2 -m_2^2 - 2m_T E_{k_2}   )  = E^*_{p_1}  E^*_{k_1}  - |\vec p^*||\vec k^*| \cos \theta^*,
\end{align}
where the incoming DM energy in the lab frame $E_{p_1}$ is known.  This allows us to obtain simple expressions for the flux factor $|\vec k^*|/|\vec p^*|$ and the scattering angle $\cos \theta^*$, giving
 \be
   d\cos \theta^* = \frac{m_T}{  |\vec p^*|  |\vec k^*| } dE_{T},
 \ee
where $E_{T} \equiv E_{k_2}$ is the energy of the recoiling target. The recoil energy distribution is
\be
\frac{d\sigma}{dE_{T}} = \frac{  m_{T} \langle |{ \cal A}|^2 \rangle}{  32 \pi  s  \left|  {\vec p}^*  \right|^{2} },
\ee
which contains the particle physics information about $d \sigma / d E_e$ needed to evaluate the signal yield in \Eq{eq:yield}. In particular, the cross section is proportional to $\epsilon^2 \alphaEM \alpha_D$.

\subsection{Numerical Signal Rate}
\label{app:NumericalRate}

Specializing to the case of elastic electron scattering $T = e^-$, $m_1 = m_2 \equiv m_\chi$, we can obtain the DM signal yield in \Eq{eq:yield} given a total production rate of
 $N_{\pizero}$ neutral pions by
\be
N_{\text{sig}} = 2 N_{\pizero} \, \Br(\pi^0 \to \chi \chi) \, n_e  \frac{1}{N_{\chi}}\sum_{i=1}^{N^c_\chi}  \ell(\vec{p}^{\,i}_\chi) \int_{E_e^{\rm low}}^{E_e^{\rm high}}dE_e \frac{d\sigma}{dE_e} (E_e, E_\chi^i)  \Theta\left[  E_\chi^i - E^{\min}_\chi(E_e)   \right].
\ee
Here $n_e$ is the target electron density, and we have used our GEANT simulation to generate a population of $N_\chi$ DM four-vectors  $\{ E_\chi, \vec{p}_\chi^{\,i} \}$. The sum is over all $N^c_\chi$ events passing geometric cuts; the path length through the detector for event $i$ is $\ell(\vec{p}^{\,i}_\chi)$, and the total geometric acceptance is $N_\chi^c/N_\chi$. To induce an electron recoil of magnitude $E_e$, the DM energy must be above the $E_{\chi}^{\rm min}(E_{e})$ threshold defined in \Eq{eq:minimumChiEnergy}.
 
For a LENA-like cylindrical detector of radius $R$ and height $h$ as discussed in \Sec{sec:Scattering}, we can compute the path length through the detector for a DM particle or neutrino.  For each geometry, we take the $z$ axis to point in the beam direction.  For the midpoint scenario depicted in \Fig{fig:LENAGeometries}a, we define the $y$ axis to be parallel to the cylindrical detector axis.  The path length is 
\be
\ell(\vec{p}_\chi) = 
\begin{cases}
S\sec \theta_y  & (\text{$\chi$ exits through side}),   \\
          (h/2 - L \tan \theta_x)\csc\theta_y  & (\text{$\chi$ exits through top/bottom}),   
\end{cases}
\ee
where $\tan \theta_{x,y} = |p_{x,y}|/p_z$  and 
\be
 S =  \frac{ D(D+2R)}{L}  - L, \qquad
 L = (R+D) \cos\theta_x -\sqrt{  (R+D)^2 \cos^2\theta_x  -  D(D  + 2 R)   }.
 \ee
 Here $L$ is the horizontal distance (parallel to the ground) $\chi$ travels prior to reaching the detector, and $D$ is the horizontal distance between the \dae\ source and the detector.
 
For the oblique scenario in \Fig{fig:LENAGeometries}b, the path lengths are 
\be
\ell(\vec{p}_\chi) = \begin{cases}
 			\left[(h + D \cos\theta_0) \tan(\theta_0  - \theta_d)  - L \right]\csc(\theta_0-\theta_d) & (\text{$\chi$ enters top/exits side}),   \\ 
                                          \left[     ( D \cos\theta_0 + h) - L  \cot(\theta_0 - \theta_d)   \right]\sec( \theta_0 -\theta_d)  & (\text{$\chi$ enters side/exits bottom}), \\ 
                                            S \csc(\theta_0 - \theta_d) & (\text{$\chi$ enters side/exits bottom}),    
\end{cases}
\ee
where $\tan \theta_0 = 2R/h$ and and $\theta_d$ is the angle with respect to the beam in the plane spanned by the beam-line and the detector's cylindrical axis.  

Finally, for the on-axis scenario in \Fig{fig:LENAGeometries}c, the cylindrical detector axis is aligned with the $z$ axis (i.e.\ the beam direction).  The path length is
\be
\ell(\vec{p}_\chi) = 
\begin{cases}
     h   \sec\theta_\chi & (\text{$\chi$ exits through bottom}), \\
     (R - D \tan \theta_\chi)\csc \theta_\chi & (\text{$\chi$ exits through side}),
     \end{cases}
 \ee
where $\tan \theta_\chi = \sqrt{p_x^2 + p_y^2} / p_z$ is the DM angle with respect to the $z$ axis. Here $D$ is the (vertical) distance between the \dae\ source and the detector.

\section{Neutrino Backgrounds}
\label{app:NeutrinoBG}

\subsection{Beam-off Backgrounds}
\label{app:NeutrinoBGBeamOff}

The irreducible background due to neutrino-electron scattering from atmospheric neutrinos is estimated from the calculated spectra of Gaisser et.\ al.~\cite{Gaisser:1988ar} with a latitude-dependent scaling factor applied to translate the flux from Kamioka to Pyhasalmi as in Ref.~\cite{LENA_DSNB}.  To determine this rate, we convolved the neutrino flux with the elastic scattering cross section. The resulting event rates were less than 1 event per year for each neutrino flavor in each energy range 106--147 MeV, 147--250 MeV, and 250--400 MeV. The CCQE scattering of atmospheric electron and muon neutrinos and antineutrinos poses an additional beam-off background. For this channel, we generated 1 million sample events on C$_{18}$H$_{30}$ using GENIE 2.8.0 \cite{Andreopoulos:2009rq}, with atmospheric flux spectra from \Ref{Battistoni:1999at} as input.  The event sample was re-weighted to match the expected number of $\nu-e$ events calculated above. After a cut requiring the outgoing lepton $\ell = e, \ \mu$ to be within $25^\circ$ of the beam direction, $\cos \theta_\ell < 0.9$ (which we take to reduce the nearly-isotropic CCQE backgrounds by a factor of 20), the raw rates for these processes are given in \Tab{tab:AllBeamOffBackgrounds}. We then assumed a 70\% reduction in the $\nu_\mu$ and $\overline{\nu}_\mu$ CCQE background rate by rejecting events followed by a Michel electron candidate, as described in \Ref{Wurm:2011zn}. Furthermore, roughly 25\% of the CCQE events for $\overline{\nu}_\mu$ and $\overline{\nu}_e$ are on hydrogen, and produce a neutron that can be tagged to reject the event; we assumed an 80\% neutron tagging efficiency. After these reductions, the dominant process in each energy range is $\nu_e$ CCQE. Using the 75\% beam-off time of \dae\ to measure this background gives a statistical uncertainty of $\sqrt{B}$ and a systematic uncertainty of $\sqrt{B/3}$, for a total uncertainty of $(\delta B)^2 = 4B/3$. We have checked that additional backgrounds such as excited resonances, coherent scattering, and deep inelastic scattering also have negligible rates compared to CCQE; in addition, these backgrounds are reducible if one can identify vertex activity or pions in the final state.

\subsection{Beam-on Backgrounds}
\label{app:NeutrinoBGBeamOn}

There are two main types of beam-on backgrounds, neutral-current elastic muon neutrino-electron scattering and CCQE neutrino-nucleon scattering. For neutrino energies $E_\nu \ll m_Z$, the differential cross section for elastic muon neutrino-electron scattering ($\nu_\mu e^- \to \nu_\mu e^-$) is 
\be
\frac{d\sigma_\nu}{dE_e} = \frac{G_F^2 s }{4\pi E_\nu} \left[  g_L^2 + g_R^2\left(1-\frac{E_e}{E_\nu} \right)^2    \right],
\ee
where $G_F$ is the Fermi constant,  $s = m_e^2 + 2 m_e E_\nu$, and $g_{L,R} = g_V \pm g_A$, where $g_V= - \frac{1}{2} + 2 \sin^2\theta_W$, $g_A = -\frac{1}{2}$.  For antineutrino scattering ($\overline{\nu}_\mu\, e^- \to \overline{\nu}_\mu\, e^-$), $g_L$ and $g_R$ are interchanged.

As outlined in \Sec{sec:BeamOnBackgrounds}, we used population of decay-in-flight pion events generated in GEANT to simulate our neutrino background events. Given a total flux $N_{\pi^+}$ of decay-in-flight $\pi^{\small +}$, each of which produces one $\nu_\mu$, the total $\nu_\mu$ background count is
\be
N_{\rm bg} =  N_{\!\pi^+} \, n_e  \frac{1}{N_{\nu}}\sum_{i=1}^{N^c_{\nu}}  \ell(\vec{p}^{\,i}_\nu) \int_{E_e^{\rm low}}^{E_e^{\rm high}}dE_e \frac{d\sigma_\nu}{dE_e} (E_e, E_\nu^i)  \Theta\left[  E_\nu^i - E^{\min}_\nu(E_e)   \right],
\ee
where as above, $n_e$ is the detector electron density, $N_\nu$ is the number of sample neutrino events generated, $N_\nu^c$ is the number of neutrino events passing geometric cuts, $\ell({\vec p}_\nu^{\,i} )$ is the path length through 
the detector for a muon-neutrino with three-momentum ${\vec p}_\nu$, and 
\be
 E_{\nu}^{\rm min}(E_{e}) =  \frac{T_e}{2} \left(
1 +  \sqrt{1 + \frac{2m_e}{T_e}}                 \,   \right),  \qquad T_e \equiv E_e - m_e,
\ee 
is the minimum neutrino energy to trigger an electron recoil of energy $E_e$. For neutrinos produced from incident $\pi^-$ or $\mu^+$, we replace $N_{\pi^+}$ by the flux of the particle in question. We have checked that the neutrino events produced by GEANT, and neutrinos obtained from manually decaying a sample of energetic pion events from GEANT, give the same results.

For the CCQE events, we used the same GENIE simulation \cite{Andreopoulos:2009rq} as for beam-off backgrounds, with an input neutrino flux spectrum generated from our GEANT simulation. We manually decayed the GEANT sample of decay-in-flight pions to obtain the spectrum for the relevant neutrino flavors, input the neutrino spectrum into GENIE with the angle-dependent path length appropriate for the geometry in question, and re-weighted the event sample to match our elastic scattering simulations. The resulting raw rates are given in \Tab{tab:AllBeamOnBackgrounds}; the same Michel electron and neutron tagging reductions apply for the background rates. We cross-checked the GENIE results by implementing the Llewellyn Smith CCQE parameterization \cite{LlewellynSmith:1971zm} in our own simulation. We find excellent agreement with GENIE, which is somewhat surprising as it implies that Pauli blocking is not significant in this energy range. We leave a detailed study of the kinematics of the CCQE background to future work. We attempted to directly simulate the background from $\mu^+$ decays with GEANT, but limited statistics proved prohibitive; as explained in the text, we expect this background to be subdominant to the other processes we have considered.

\bibliography{main.bib}

\end{document}